\documentclass[trackchanges,twocolumn]{aastex701}

\usepackage{bm}

\begin{document}

\title{Solar Energetic Particle Reflection by Precursor ICMEs:\ Multi-spacecraft Observations of Bi-Directional Electron Beams at 1 AU}

\author[orcid=0000-0002-4820-8594]{Lucas Liuzzo}
\affiliation{Space Sciences Laboratory, University of California, Berkeley}
\email{liuzzo@berkeley.edu}

\author[orcid=0000-0001-8495-9179]{Wenwen Wei} 
\affiliation{Space Sciences Laboratory, University of California, Berkeley}
\email{wwwei@berkeley.edu}

\author[orcid=0000-0001-8137-8176, sname='Poppe']{Andrew R. Poppe}
\affiliation{Space Sciences Laboratory, University of California, Berkeley}
\email{poppe@berkeley.edu}

\author[orcid=0000-0002-1604-3326, sname='Lee']{Christina O. Lee}
\affiliation{Space Sciences Laboratory, University of California, Berkeley}
\email{clee@ssl.berkeley.edu}

\author[orcid=0000-0001-7024-1561]{Vassilis Angelopoulos}
\affiliation{Department of Earth and Space
Sciences, University of California, Los Angeles}
\email{vassilis@ucla.edu}

\begin{abstract}

We present case studies of two impulsive solar energetic electron (SEE) events during which particles at energies from 1--600 keV were detected by THEMIS-ARTEMIS orbiting the Moon, Wind at Earth's first Lagrange point, and (for one event) STEREO-A located at 1 AU, off the Sun-Earth line.
The SEEs were initially highly anisotropic, traveling outward along the magnetic field with distinct energy-time dispersion.
For one event, the spectra contained inverse velocity dispersion (IVD) signatures, whereby electrons at intermediate energies arrived to the spacecraft before those at higher energies.
Similar features were recently discovered within 1 AU for energetic protons; this represents the first IVD detection for energetic electrons at Earth's orbital distance.
During both events, a second beam of counter-streaming electrons was detected after a short time.
Based on the time-delay in the detections at various energies, the path traveled by these counter-streaming electrons was on the order of 1--2 AU.
We show that an interplanetary coronal mass ejection (ICME) passed the spacecraft a few days prior to the onset of each event and was located beyond 1 AU when the SEEs were detected, suggesting that the electrons were part of the same population, but reflected off the shock front of these precursor ICMEs.
In the context of solar system exploration, this represents an unidentified hazard for astronaut safety beyond low-Earth orbit: although the initial phase of impulsive SEE events typically stream anti-Sunward, ICMEs located beyond Earth provide a mechanism for hazardous particles to travel Sunward during extreme events.

\end{abstract}



\section{Introduction}
Impulsive solar energetic electrons, typically accelerated by high-energy solar flares (e.g., \citealp{Lin1985,Reames1999,Krucker2007}), travel rapidly outward through the solar system along the interplanetary magnetic field (IMF).
Due to their high velocities and the decreasing IMF magnitude as a function of radial distance, the pitch angle distributions (PADs) of these particles are typically highly beamed by the time they reach a radial distance of 1 AU.
Solar energetic electron (SEE) observations also often display a time dispersion as a function of energy, whereby the highest energy particles (with largest velocities) arrive at 1 AU before their lower-energy counterparts (see, e.g., \citealp{Lin1974,Krucker1999}).
Deviations from these typical signatures can offer important diagnostic information and insight into more complex transport and generation mechanisms for solar energetic particles throughout the solar system.

\begin{figure*}[!t]
    \centering
    \includegraphics[width=0.95\textwidth]{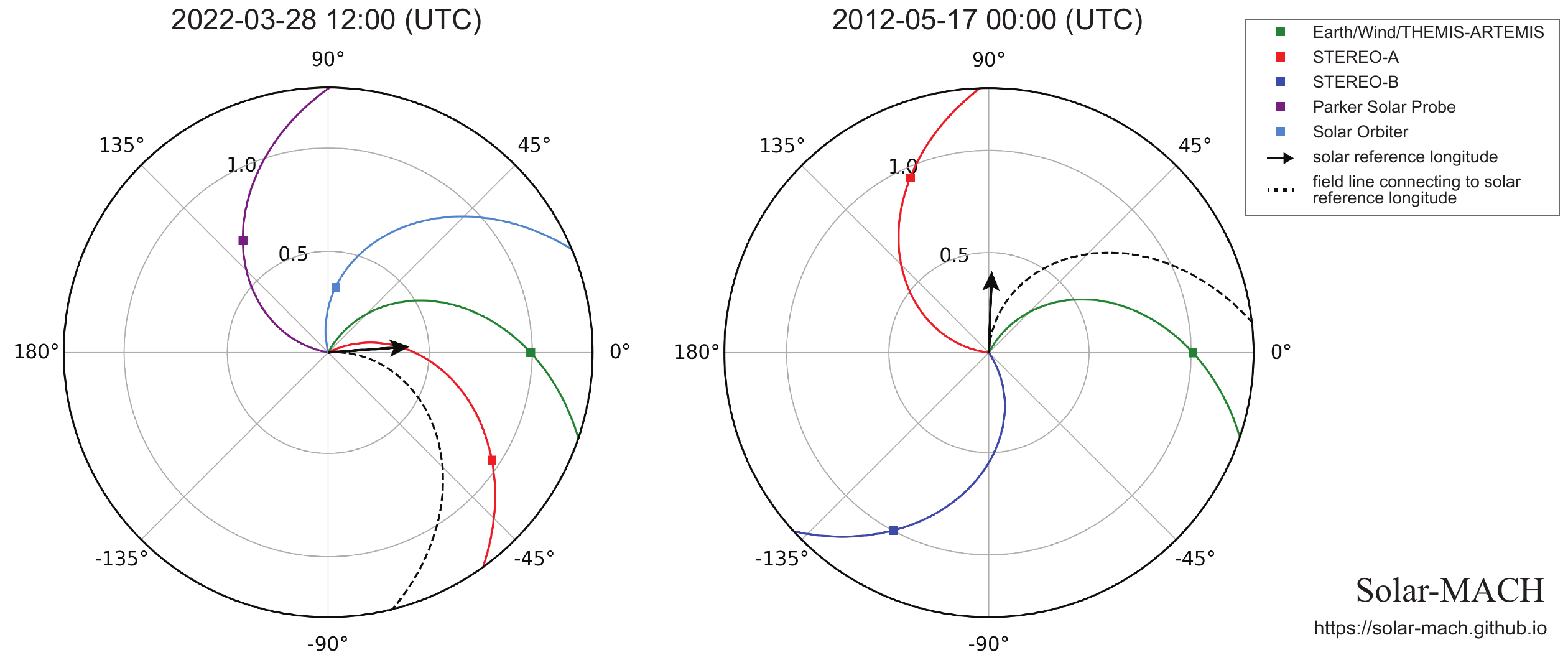}
    \caption{
        Radial and azimuthal distribution of Earth and multiple spacecraft throughout the solar system plotted in the Stonyhurst Heliographic coordinate system during the (left) 28 March 2022 and (right) 17 May 2012 events.
        The black arrows denote the reference longitude of the flares that generated the observed SEEs.
        Included Parker spiral field lines assume a solar wind velocity of 400 km/s.
        }
    \label{fig:WindArtSta}
\end{figure*}

One such deviation that has recently been identified in the transport of solar energetic \textit{protons} observed in the inner heliosphere within 1 AU has been termed ``inverse velocity dispersion'' (IVD).
This phenomenon is characterized by protons at lower energies arriving earlier than those at higher energies (see \citealp{Cohen2024,Ding2025a}), but any detection of such signatures for energetic \textit{electrons} has not been made.
Additionally, deviations to the particle PAD from uni-directional anisotropic beams can reveal information about the underlying path the particles traveled before their detection.
For example, observations of counter-streaming SEEs can indicate trapping within a large-scale flux rope structure propagating outward through the solar system \citep{Lario2009}, or even particles mirroring off of a boundary located elsewhere in the solar system before returning to the observer \citep{Ding2025}.
However, clear multi-point observations of counter-streaming SEE events are scarce, limiting our ability to better contextualize these events \citep{Tan2009}.

In this study, we present multi-spacecraft observations of two impulsive, bi-directional SEE events associated with M-class solar flares.
These electrons span energies ranging from approximately 1--600 keV, and were detected by the Wind spacecraft \citep{Acuna1995}, the two probes of the \textit{Time History of Events and Macroscale Interactions during Substorms}-\textit{Acceleration, Reconnection, Turbulence, and Electrodynamics of the Moon's Interaction with the Sun} (THEMIS-ARTEMIS) mission \citep{Angelopoulos2008,Angelopoulos2011}, and for one event, by the \textit{Solar TErrestrial RElations Observatory-Ahead} (STEREO-A) spacecraft \citep{Kaiser2008}.
At onset of one of these events, we report the first detection of IVD signatures for energetic electrons along Earth's orbit.
We also provide evidence supporting the observation of counter-streaming SEEs that reflected off of the shock front of ICMEs that had passed the spacecraft multiple days prior to each of the events' onsets, and were located approximately 1--2 AU beyond Earth's orbit at the time of the events.
These results provide important context for future crewed exploration beyond low-Earth orbit and highlight the importance of accurately characterizing the role that past space weather conditions have on shaping properties of ongoing energetic particle events.

\section{Methods and Observational Data}
For this analysis, we use data from the magnetic field and particle instruments on multiple spacecraft located throughout the solar system.
To help constrain the ambient plasma conditions before reaching Earth, we use data from the Wind spacecraft \citep{Harten1995}, which has been located at the Sun-Earth first Lagrange point since 2004, located nearly $200$ Earth radii upstream of the planet.
Onboard, the \textit{Three-Dimensional Plasma} instrument (3DP; \citealp{Lin1995}) detects measurements of electrons spanning energies from $\sim5$ eV up to approximately $1$ MeV, representing a near-continuous data stream of the solar wind and energetic electron conditions incident onto Earth.
Next, to quantify the plasma environment near the Moon, we apply data from the THEMIS-ARTEMIS mission \citep{Angelopoulos2008,Angelopoulos2011}, which has been orbiting the Moon since 2011.
The two probes of this mission (P1 and P2) are identically instrumented, with \textit{Electrostatic Analyzers} (ESA) and \textit{Solid State Telescopes} (SST) that share heritage with 3DP on Wind, and detect electrons from energies $5$ eV $\le E \le 1$ MeV (e.g., \citealp{McFadden2008a}).
{Finally}, we use data from STEREO, which consist of two spacecraft (Ahead and Behind) with identical instrumentation.
Using the \textit{Solar Wind Plasma Electron Analyzer} 
and \textit{Solar Electron Proton Telescope} (SEPT) instruments \citep{STEREOSWEA,STEREOSEPT}, the probes detect electrons from 1 eV to 3 keV, and 20 keV to 400 keV, respectively.
Contact was lost with STEREO-B in 2016, but STEREO-A continues to operate.

We focus on two separate energetic particle events for our study.
Figure \ref{fig:WindArtSta} displays the directionality of the flare during each of the events (i.e., the solar reference longitude included in the panels), with the positions of Earth and various spacecraft shown.
The first SEE event we study was associated with an M4.0 flare that erupted from active region AR12975 (N14, W04) on 28 March 2022 with a peak X-ray flux occurring at 11:29 UTC.
Wind, STEREO-A, Solar Orbiter (SolO), and Parker Solar Probe (PSP) all detected a type III radio burst during the event.
Wind, and THEMIS-ARTEMIS P1 and P2, located near the Sun-Earth line, observed an impulsive enhancement in the energetic electron differential flux approximately 30 minutes after the peak of this flare.
STEREO-A, which was located nearly $30^\circ$ azimuthally off of the Sun-Earth line (see Figure \ref{fig:WindArtSta}), also detected an impulsive energetic electron event.
While the SolO \textit{Spectrometer Telescope for Imaging X-rays} and the \textit{Extreme Ultraviolet Imager} instruments observed the flare (see \citealp{Purkhart2024}), electrons from the associated solar energetic particle event were not detected.
Data from the \textit{Integrated Science Investigation of the Sun} energetic particle instrument on PSP \citep{McComas2016} from this time did not show any signatures of energetic electrons.
For this event, we therefore focus on the energetic electron data collected by Wind, STEREO-A, and THEMIS-ARTEMIS P1 and P2.

The second event that we present occurred on 17 May 2012 and was associated with a M5.1 solar flare that erupted from active region 1476 (N07, W88) with a peak X-ray flux occurring at 01:47 UTC.
Associated with this flare was a high-speed (exceeding $1500$ km/s) interplanetary coronal mass ejection (ICME) that hit STEREO-A approximately one day later.
Various Earth-based spacecraft detected the first SEEs approximately 15 minutes after this peak, and the subsequent energetic particle storm resulted in the first ground level enhancement of solar cycle 24 \citep{Gopalswamy2013,PerezPeraza2018}.
The event was widespread throughout the inner heliosphere, with STEREO-A and STEREO-B each measuring enhanced fluxes from this event, but detecting only gradual increases in the solar energetic electron flux over a few days compared to the more impulsive observations seen near Earth (see, e.g., \citealp{Dresing2014}).
For this reason, we focus on presenting the energetic electron data collected by Wind and the THEMIS-ARTEMIS probes (P1 and P2) located along the Sun-Earth line.

\section{Results}
\subsection{The 28 March 2022 Event}

\begin{figure}[!t]
    \centering
    \includegraphics[width=0.47\textwidth]{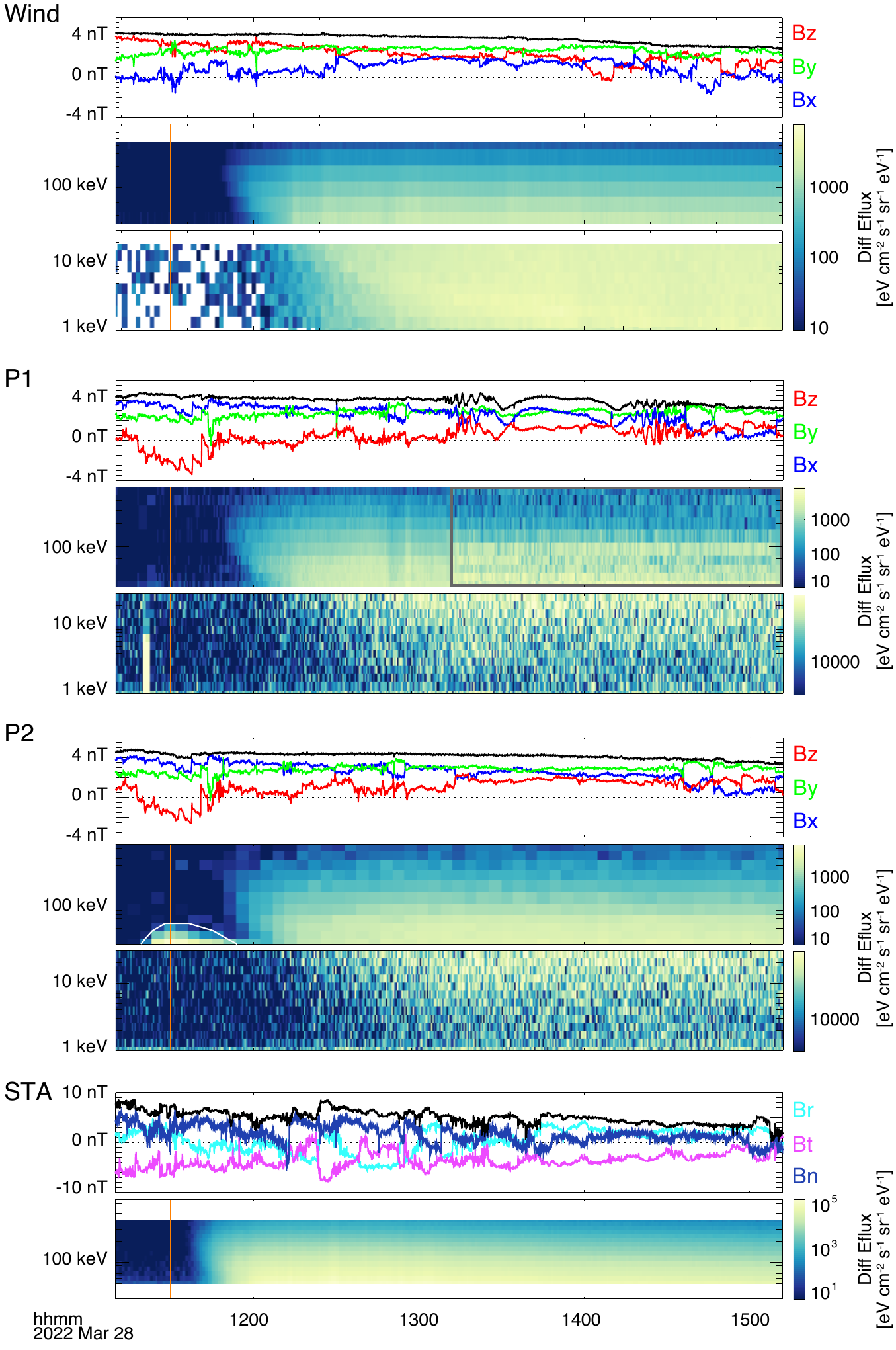}
    \caption{
        Observations of a solar energetic electron event on 28 March 2022 by Wind, THEMIS-ARTEMIS P1, P2, and STEREO-A.
        The magnetic field is displayed in the GSE coordinate system for Wind, P1, and P2, and the RTN coordinate system for STEREO-A.
        The differential energy fluxes are omnidirectional.
        To guide the eye, the vertical orange line denotes the time of the flare's peak X-ray flux, and the white semi-circle in the P2 SST channel denotes contamination in the sensor from these X-rays.
        The gray box in the panel displaying the P1 fluxes corresponds to times when the SST attenuator was activated, where a correction to the omnidirectional fluxes has been applied.
        }
    \label{fig:fluxes2022}
\end{figure}

Figure \ref{fig:fluxes2022} displays observations from the solar energetic electron event on 28 March 2022, grouped into observations by the (from top to bottom) Wind, THEMIS-ARTEMIS P1, THEMIS-ARTEMIS P2, and STEREO-A spacecraft.
During the entirety of this event, P1 and P2 were located upstream of Earth in the solar wind.
The top panel of each grouping displays the measured magnetic field.
For Wind, P1, and P2, this is plotted in the GSE coordinate system, in which unit vector $\bm{\hat{\mathrm{x}}}$ points toward the Sun, $\bm{\hat{\mathrm{z}}}$ is aligned with the ecliptic north pole, and $\bm{\hat{\mathrm{y}}}$ completes the right-handed set.
For STEREO, the magnetic field is shown in the RTN system, where $\bm{\hat{\mathrm{R}}}$ is the line extending radially away from the Sun and connecting through the spacecraft, $\bm{\hat{\mathrm{T}}}$ is the cross product of the Solar rotation vector with $\bm{\hat{\mathrm{R}}}$, and $\bm{\hat{\mathrm{N}}}$ completes the right-handed system.

For each grouping of spacecraft observations, the panels following the magnetic field display the electron differential energy fluxes.
For Wind and the two THEMIS-ARTEMIS probes, the fluxes as measured by the SSTs ($E \gtrsim 30$ keV) and ESAs ($E \lesssim 30$ keV) are shown, while for STEREO-A the SEPT measurements ($50$ keV $\lesssim E \lesssim 400$ keV) are included.
The vertical orange line overlaid on each panel denotes the timing of the peak X-ray flux output by the M-class flare on this date.
Note that for this event, X-rays associated with this flare were detected by the SST onboard THEMIS-ARTEMIS P2, which generates secondary energetic electrons in the instrument (e.g., \citealp{Larson2015}).
Hence, the enhancement in the P2 $E \le 60$ keV channels (bounded by the white line included in the P2 SST panel in Figure \ref{fig:fluxes2022}) is not caused by solar energetic electrons but instead represents contamination in the SST during this event.
In addition, the SST attenuator on P1 activated just after 13:00 UTC (denoted by the gray box in Figure \ref{fig:fluxes2022}).
This is designed to protect the instrument from intense moonlight near periselene and lasted for the entire passage through the lunar wake, which is visible in the magnetic field near 13:45 UTC.
However, this attenuation results in poor counting statistics in the SST due to the associated reduction in the instrument's geometric factor, rendering measurements from that time unreliable.
For the omnidirectional fluxes included in Figure \ref{fig:fluxes2022}, we have applied a correction factor to counter this attenuation where required, but the absolute values during this time period should be interpreted with caution.

As visible in each panel, the first SEEs to arrive at each of the four spacecraft are detected anywhere from $\sim10$ min (STEREO-A) to $\sim20$ min (Wind, P1, P2) after the flare.
Flare-accelerated electrons typically display a velocity dispersion signature: since electrons with higher energies travel faster, they tend to arrive at a detector near 1 AU earlier than electrons at lower energies (e.g., \citealp{Reames2021})
However, for the electrons detected by each of the near-Earth spacecraft shown in Figure \ref{fig:fluxes2022}, the differential fluxes have a ``nose energy'' near $E \approx 300$ keV.
That is, $E \approx 300$ keV electrons are the \textit{first} electrons that arrive at Wind, P1, and P2.
While an (expected) dispersive signature is visible at lower energies (extending into the ESA energy channels for P1, P2, and Wind), an inverse velocity dispersion is seen at higher energies: the differential fluxes of these highest-energy electrons increases only after the initial $E \approx 300$ keV electrons arrive.

\begin{figure*}[!t]
    \centering
    \includegraphics[width=0.95\textwidth]{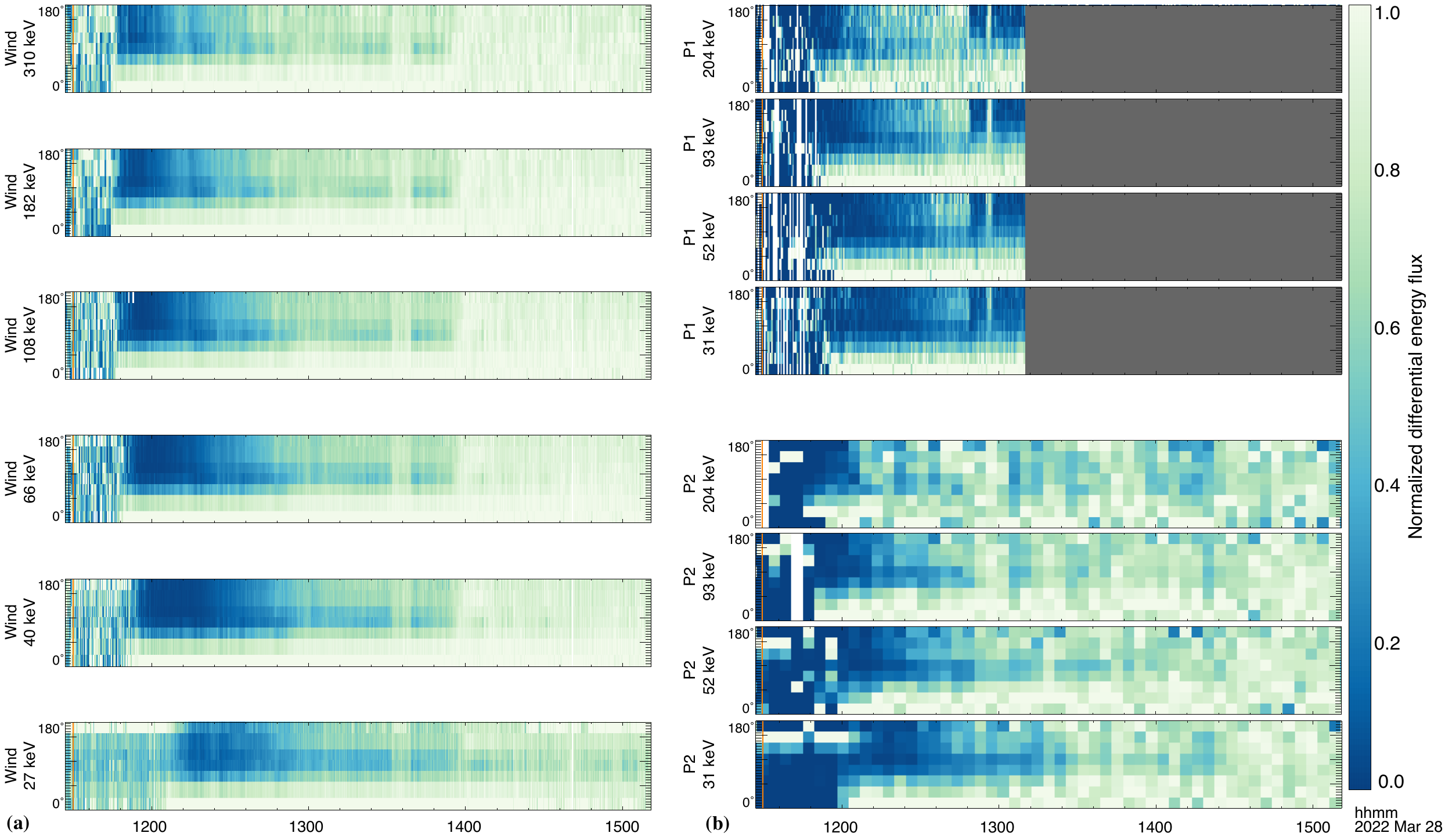}
    \caption{
       Normalized energetic electron pitch angle distributions observed by (a) Wind and (b) THEMIS-ARTEMIS at select energies during the 28 March 2022 event.
       Gray shaded regions within the P1 panels correspond to times when the SST attenuator was activated.
       The vertical orange line denotes the time of the flare's peak X-ray flux.
        }
    \label{fig:pas_2022}
\end{figure*}

Such IVD signatures have recently been reported for multiple solar energetic proton events detected by PSP.
One such event associated with a CME occurred in September 2022, where \cite{Cohen2024} present detections of protons with a nose energy near $E \approx 1$ MeV when the spacecraft was located approximately 15 solar radii from the Sun.
Below this energy, typical velocity dispersion was measured, but above this energy a similar inverse signature was detected.
\cite{Cohen2024} suggest that the responsible acceleration mechanism requires less time to accelerate protons to lower energies than it does to higher energies.
Hence, the lower energy protons are ``released'' earlier---and detected sooner---than their higher energy counterparts.
Additional modeling by \cite{Ding2025a} has further supported the idea that the maximum energy to which protons are accelerated along a shock front increases with time.
Hence, although the higher-energy particles have larger velocities, they are generated later than their lower-energy counterparts and do not have sufficient time (or distance) to overtake the lower energy protons, thereby causing an IVD signature (see especially their Figures 6--8).
In addition, \cite{Ding2025a} show that solar energetic proton IVD signatures observed by SolO may also be affected by the spacecraft's evolving magnetic connectivity to the shock during its propagation.
Finally, after surveying energetic ion data from STEREO, that study also reported multiple, previously unreported observations of IVD signatures in SEPT data.
Additional IVD signatures in energetic proton data have recently been identified in SolO observations as well \citep{Li2025}.

However, this is the first report of such IVD signatures for solar energetic \textit{electrons}.
Although electrons of a given energy travel much faster than protons, the much larger distance of the spacecraft from the Sun for the observations shown in Figure \ref{fig:fluxes2022} compared to those by PSP or SolO (near 1 AU, compared to $0.07$ AU for PSP or $0.5$ AU for SolO) suggests that the mechanism that generated the IVD signature during this event may have been different.
Importantly, although the IVD signature was similar between the Wind and two THEMIS-ARTEMIS spacecraft, Figure \ref{fig:fluxes2022} shows that it was substantially less pronounced at STEREO-A.
This may suggest differences between how the spacecraft were connected to the flare region, or perhaps some time dependence in the electrons' acceleration mechanism that was more prolonged along the Sun-Earth line during this specific SEE event.
The exact generation process behind the electron IVD signature (or indeed, how common such signatures are) is an open question and lies beyond the scope of this current study.

Figure \ref{fig:pas_2022} displays the PADs of electrons at select energies for (left) Wind and (right) THEMIS-ARTEMIS during this event.
For each point in time, the fluxes at a given energy are normalized to the maximum value detected across all pitch angles.
Figure \ref{fig:pas_2022} illustrates that starting between approximately 11:48 UTC and 12:00 UTC, the three spacecraft detected a highly anisotropic beam of solar energetic electrons with pitch angles near $0^\circ$--$45^\circ$.
As in Figure \ref{fig:fluxes2022}, the arrival time of these electrons below the nose energy of $E \lesssim 300$ keV were dispersive in energy, with the highest energy electrons measured by the spacecraft arriving up to 12 minutes sooner than the lowest energies detectable by the SSTs.
For Wind and the THEMIS-ARTEMIS probes, the IMF during this time was directed along the GSE $+\bm{\hat{x}}$ and $+\bm{\hat{y}}$ directions (see Figure \ref{fig:fluxes2022}).
This corresponds to a local magnetic field that was nearly perpendicular to the nominal Parker spiral direction(associated with a magnetic field pointing along either $+\bm{\hat{x}}$ and $-\bm{\hat{y}}$ or $-\bm{\hat{x}}$ and $+\bm{\hat{y}}$) and indicates that a kink or reversal in the orientation of the IMF was detected by these Earth-based spacecraft.
Although these SEEs were generated by a solar flare and traveled in an outward direction along the solar wind magnetic field lines, the orientation of the IMF near Wind and THEMIS-ARTEMIS means they were locally detected traveling toward the Sun at this time.

At each energy, this strong collimated beam of field-aligned electrons was observed for $\sim20$ min.
After this period, a second beam of anti-field-aligned electrons (pitch angles near $180^\circ$) were detected.
The normalized flux of these electrons never equals that of the field-aligned electrons at any energy, remaining below a value of approximately $0.8$ during the three-hour period shown in Figure \ref{fig:pas_2022}.
As with the field-aligned electrons, the arrival times of the anti-field-aligned beams were also dispersive in energy, beginning as soon as 12:10 UTC for the highest energy channels and as late as 13:00 UTC for the lowest energies.
In P1, the anti-field-aligned electron beam was interrupted at all energies near 13:00 UTC, associated with a rotation of the local magnetic field (see Figures \ref{fig:fluxes2022} and \ref{fig:pas_2022}).
It is possible that the rotation caused the field line to intersect the Moon, causing these electrons to be ``shadowed'' by the lunar surface (see also \citealp{Liuzzo2024SEP}).
This mechanism is further supported by the observations in P2, which detected a similar magnetic field rotation at this time but no change in the electron PADs across all energies.
Notably, a band of minimum flux at pitch angles near $90^\circ$ is visible for nearly two hours across all energies at each of the spacecraft.
In some energy channels, this band extends for even longer times; see, e.g., the $E = 27$ keV channel on Wind and $E = 31$ keV channel on P2.
We note that this is \textit{not} a reduction in the flux of electrons at near-perpendicular pitch angles.
Instead, since the fluxes in Figure \ref{fig:pas_2022} are normalized, the enhanced fluxes of the field-aligned and anti-field-aligned electrons causes the normalized flux near more perpendicular pitch angles to appear as minima in the PADs.

\begin{figure}[!t]
    \centering
    \includegraphics[width=0.47\textwidth]{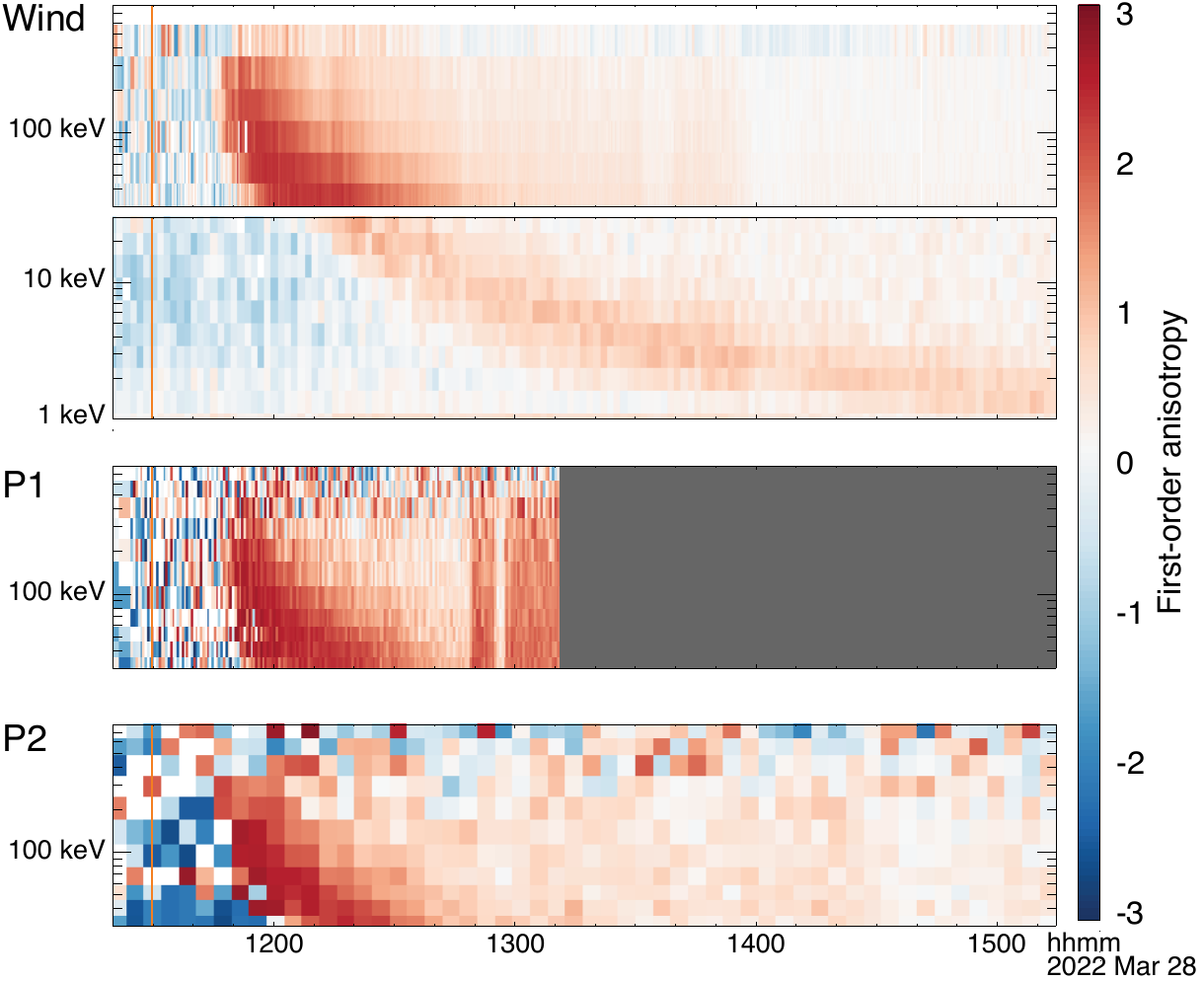}
    \caption{
       First-order anisotropy of the energetic electrons observed by Wind and THEMIS-ARTEMIS P1 and P2 during the 28 March 2022 event.
       The gray shading and orange vertical line are as in previous figures.
        }
    \label{fig:anisotropies}
\end{figure}

To illustrate this in a different way, Figure \ref{fig:anisotropies} displays the first-order anisotropy of the SEEs detected by Wind, P1, and P2.
For each energy, this quantity is calculated by
\begin{equation}
    A = 3 \frac{\int_{-1}^{1} I(\mu) \mu d\mu}{\int_{-1}^{1} I(\mu) d\mu} \quad,
\end{equation}
where $I(\mu)$ is the pitch-angle dependent electron differential energy flux and $\mu$ is the cosine of the average pitch angle (see also, e.g., \citealp{Tan2007,Wei2024,Ding2025}).
For electrons originating from the Sun, the quantity $A$ is maximized when $\mu$ is positive (i.e., for a fully field-aligned electron population; red coloring in Figure \ref{fig:anisotropies}), and minimized when $\mu$ is negative (for a fully anti-field-aligned population; blue coloring in Figure \ref{fig:anisotropies}).
Hence, the initial arrival of the field-aligned electron beam results in a strong (positive) first-order anisotropy in the electron distributions, which gradually reduces to near zero as the anti-field-aligned, counter-streaming electrons are detected.
Since the normalized flux of the anti-field-aligned electrons remains below that of the initial field-aligned electrons, the sign of the first-order anisotropy stays positive.
In the panel for P1, the rotation in the magnetic field near 13:00 UTC that caused the anti-field-aligned electron beam to become shadowed by the lunar surface and disappear is immediately visible, where the first-order anisotropy suddenly increases.

Two potential mechanisms may explain the initial observations of dispersive, field-aligned electron beams, separated in time by observations of dispersive, anti-field aligned electrons.
The first posits that these electrons represent two separate populations, each with a unique generation mechanism, that approach the spacecraft along the local magnetic field from opposite sides (i.e., traveling within a closed magnetic loop).
The delay between the detection of the two populations could be attributed to either a time-delay in distinct acceleration processes (e.g., two separate flares occurring at different time or a flare with an associated CME accelerating two different populations), or to a similar acceleration mechanism, but traveling along distinct paths of disparate lengths.
Such signatures have been observed at Earth and elsewhere in the solar system, alike (e.g., \citealp{Miroshnichenko2005,Ruffolo2006,Saiz2008,Ding2025}).
The second potential mechanism hypothesizes that these electrons are part of the \textit{same} population but are observed multiple times: electrons streaming outward from the Sun would be initially detected, followed by a second detection of inward-streaming electrons.
In this case, the shape of the electron spectrum should be similar, since they represent multiple observations of the same population.
Such a mechanism would point to a boundary located beyond the observer off of which the electrons could reflect and return.
In this scenario, the initial dispersion that was observed in the onset time between electrons at the highest and lowest energies should continue to grow due to their differential velocities, and become even more pronounced during subsequent detections.
A similar mechanism has been reported for Wind observations of bi-directional energetic electrons, where anti-Sunward-traveling particles were accelerated at the shock front of an ICME located close to the Sun, and Sunward-traveling particles were reflected off of a shock located beyond 1 AU \citep{Li2020}.

\begin{figure*}[t!]
    \centering
    \includegraphics[width=0.95\textwidth]{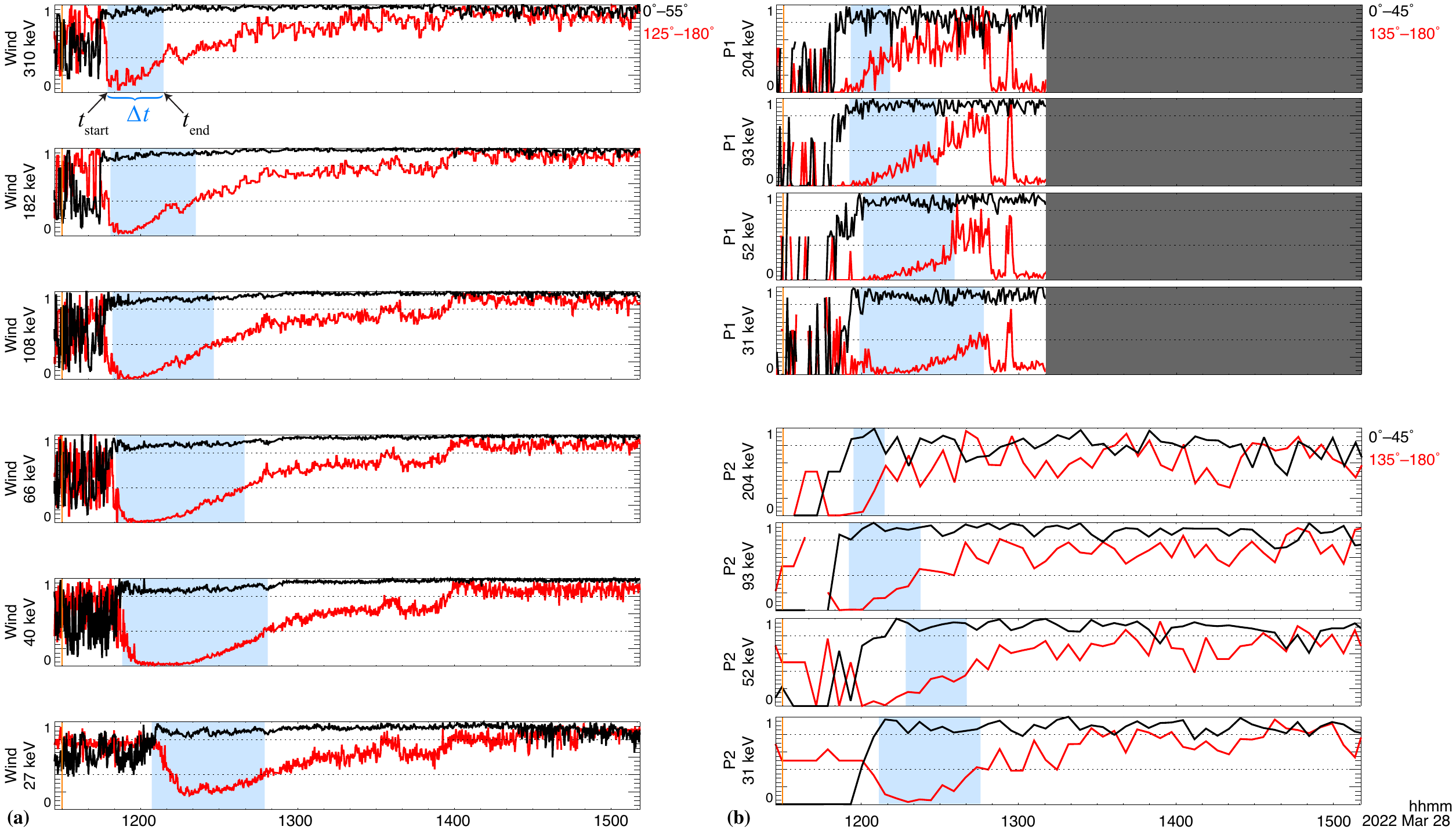}
    \caption{
       Timeseries of normalized electron fluxes at select energies observed by (a) Wind and (b) THEMIS-ARTEMIS P1 and P2 during the 28 March 2022 event.
       Black or red lines correspond to electrons with field-aligned or anti-aligned pitch angles, respectively.
       Blue shaded regions denote the time $\Delta t$ between detections of initial and counter-streaming electrons.
       The gray shading and orange vertical line are as in previous figures.
       }
    \label{fig:windartlines2022}
\end{figure*}

To help illuminate which mechanism could explain these observations during the SEE event on 28 March 2022, Figure \ref{fig:windartlines2022} shows the timeseries of (normalized) electron fluxes observed by Wind and THEMIS-ARTEMIS P1 and P2 at select energies of their respective SST instruments.
For each energy, black curves indicate the field-aligned population (with pitch angles 0$^\circ$--55$^\circ$ for Wind and 0$^\circ$--45$^\circ$ for THEMIS-ARTEMIS), while red curves indicate the anti-field-aligned population (pitch angles 125$^\circ$--180$^\circ$ for Wind and 135$^\circ$--180$^\circ$ for THEMIS-ARTEMIS).
The blue shaded region in each panel indicates the time difference, $\Delta t = t_\mathrm{end} - t_\mathrm{start}$, between the detection of these two oppositely traveling electron populations.
For each energy, we define $t_\mathrm{start}$ as the time where the ten-minute running average for the normalized flux of the \textit{field-aligned} population exceeds a value of 0.8, and $t_\mathrm{end}$ is when the ten-minute running average for the flux of the \textit{anti-field-aligned} population exceeds a value of 0.4.
Note that the blue bars therefore do not exactly align with the locations where the black and red curves cross the 0.8 and 0.4 thresholds, respectively, since these curves show the instantaneous values (ignoring the instrument noise contribution to each measurement).

As visible in Figure \ref{fig:windartlines2022}, the values of $t_\mathrm{start}$ and $t_\mathrm{end}$ grow with decreasing energy for the Wind and THEMIS-ARTEMIS observations.
Consistent with Figures \ref{fig:fluxes2022} and \ref{fig:pas_2022}, this illustrates dispersion in the SEE population.
In addition, the values of $\Delta t$ generally grow with {decreasing} energy; i.e., the dispersion in the anti-field-aligned electrons is more exaggerated with decreasing energy compared to their field-aligned counterparts.
This behavior is clearest in the Wind observations, with a similar behavior in THEMIS-ARTEMIS P1.
However, the low temporal resolution of the P2 SST during this time makes it difficult to identify a similar feature in this instrument.
Regardless, this behavior is consistent with the SEEs belonging to the same population, where the initially field-aligned electrons travel past the spacecraft, reflect off of a boundary, and return with an anti-field-aligned component.

\begin{figure}[t!]
    \centering
    \includegraphics[width=0.45\textwidth]{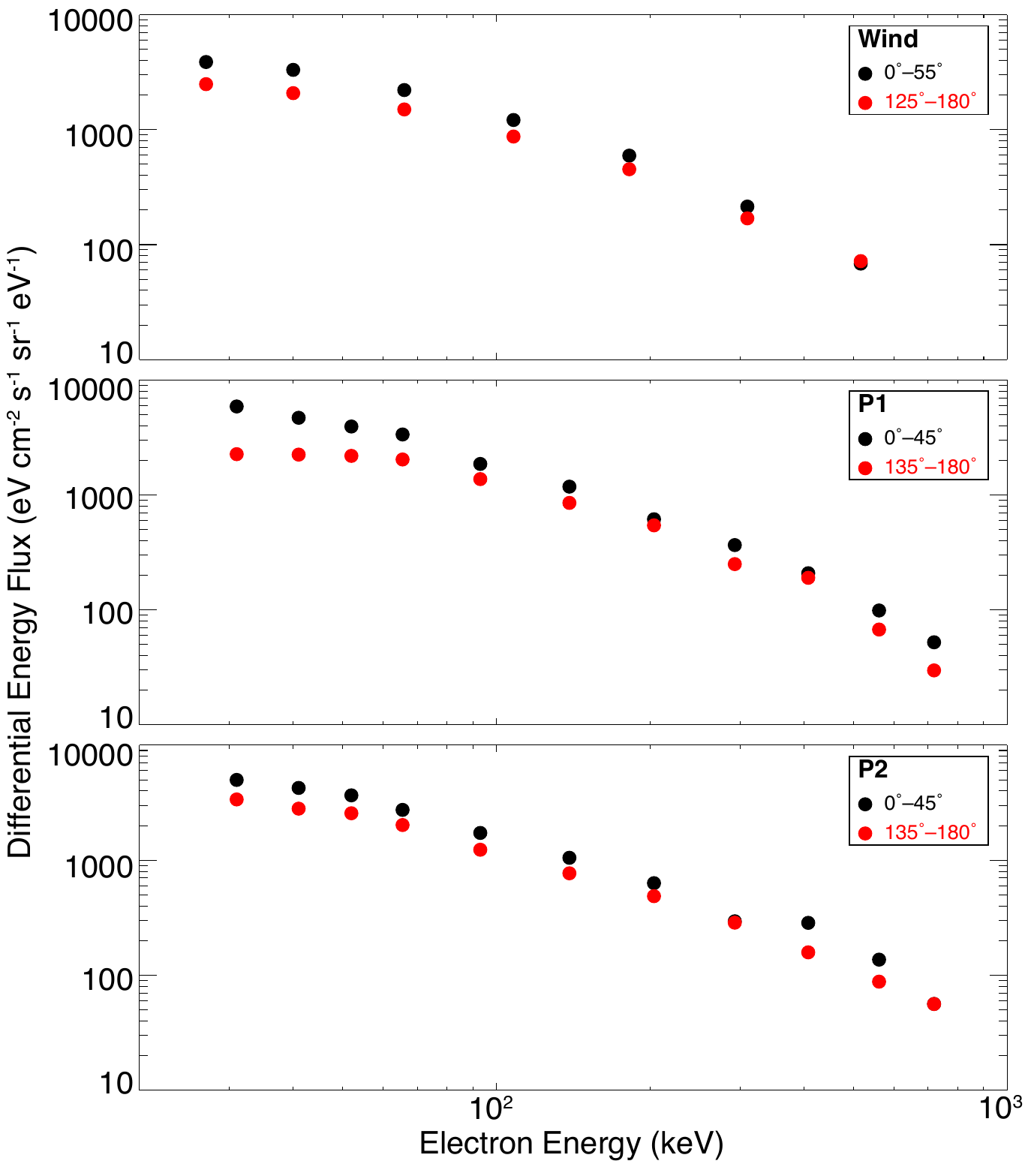}
    \caption{
    Solar energetic electron spectra during the 28 March 2022 event observed by (top) Wind, THEMIS-ARTEMIS (middle) P1, and (bottom) P2.
    Spectra are shown separately for the (black) field-aligned and (red) anti-field-aligned electrons.
    See text for further detail.    
    }
    \label{fig:spectra}
\end{figure}

To provide further evidence that these electrons are part of the same population, Figure \ref{fig:spectra} displays the spectra of the solar energetic electrons during the 28 March 2022 event, again separated into (black) field-aligned and (red) anti-field-aligned pitch angles.
For each spacecraft, these spectra are averaged over a time when both field-aligned and anti-field-aligned electrons were detected.
For Wind and THEMIS-ARTEMIS P2, this corresponds to a one-hour average between 13:50 UTC and 14:50 UTC.
However for THEMIS-ARTEMIS P1, the spectrum is averaged over only ten minutes (from 12:38 UTC to 12:48 UTC), since the magnetic field rotation near 13:00 UTC and activation of the attenuator near 13:10 prevents using a wider time range.
Indeed, across nearly all energies for the three spacecraft, the spectral slopes for the field-aligned and anti-field-aligned electrons are nearly identical, with the fluxes of the two populations remaining within a factor of $\sim1.5$.
The biggest discrepancy occurs in P1 at lowest energies (below $E\approx 50$ keV), which indicates that not all reflected electrons had returned to the spacecraft yet, due to the earlier time over which the spectrum was averaged compared to the other two spacecraft.
The similarities between the spectra provide confidence that these two counter-streaming electron beams are the same population, in agreement with the hypothesis of anti-field-aligned electrons reflecting off of a magnetic boundary located elsewhere in the solar system.

\begin{figure*}[!t]
\centering
    \includegraphics[width=0.95\textwidth]{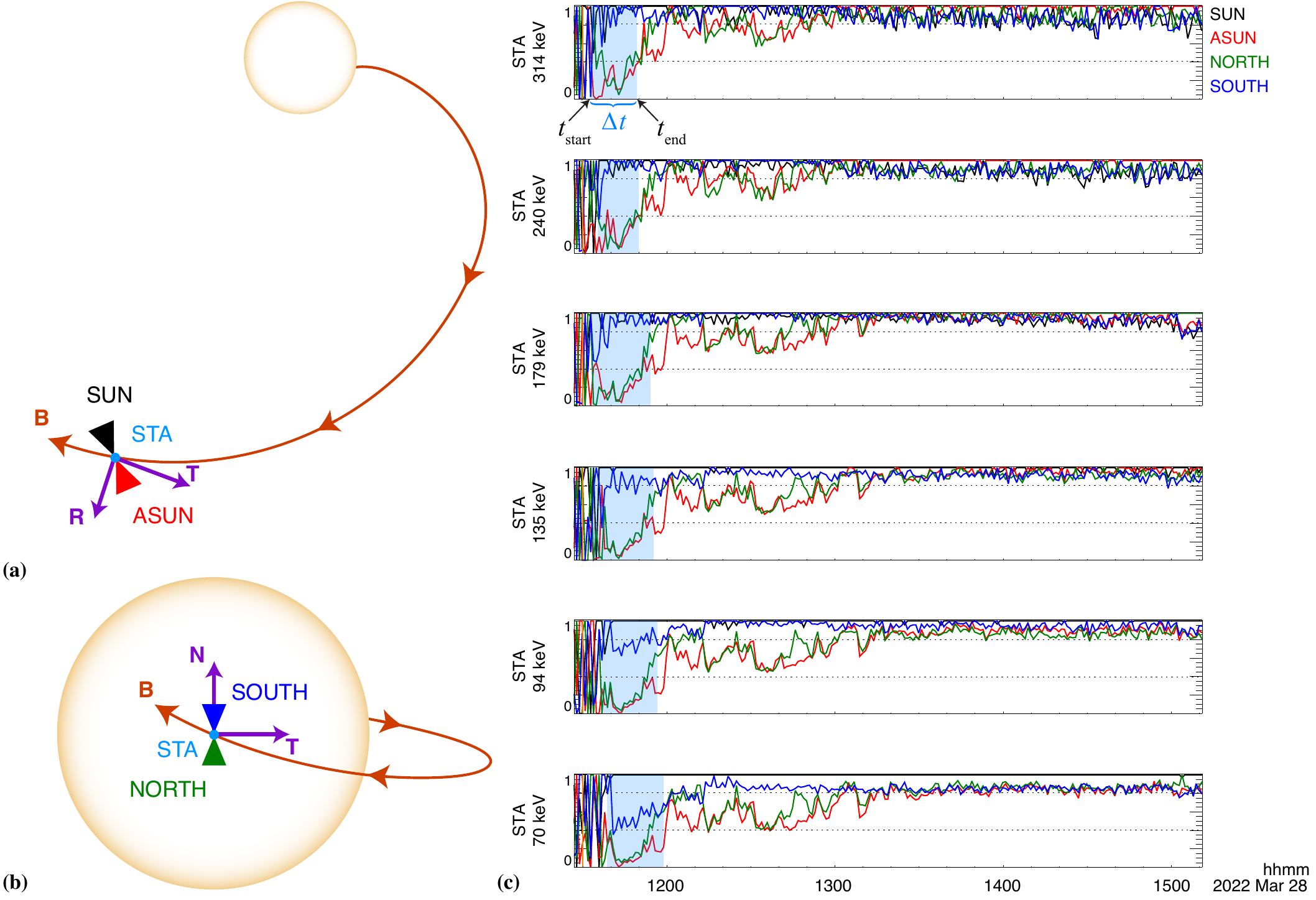}
    \caption{
      Panels (a) and (b) display a schematic demonstrating the geometry of the STEREO-A SEE observations on 28 March 2022 in two planes of the RTN coordinate system: (a) the R-T plane and (b) the T-N plane.
      The four SEPT sensors (SUN, ASUN, NORTH, SOUTH), along with a simplified, example magnetic field line connecting the spacecraft to the flare, are projected onto these planes.
      The panels are not to scale.
      Panel (c) displays normalized electron fluxes at select energies observed by the four SEPT detectors.
      Black, red, green, and blue lines correspond to electrons detected in the SUN, ASUN, NORTH, and SOUTH detectors, respectively.
      The vertical orange line and blue shaded regions are as in Figure \ref{fig:windartlines2022}.
      }
    \label{fig:stalines2022}
\end{figure*}

Figure \ref{fig:stalines2022} illustrates the timeseries of normalized electron flux detected by STEREO-A in the same style as Figure \ref{fig:windartlines2022}.
This figure also includes the viewing geometries of the four SEPT detectors during the event.
Since these detectors are fixed in space, we define the value of $\Delta t$ as the difference between the time when the 10-minute running average of \textit{any} detector has exceeded $80\%$ of the normalized flux and the time when \textit{all} detectors have exceeded $40\%$ of the flux (see Figure \ref{fig:stalines2022}c).
As with the Wind and THEMIS-ARTEMIS observations, the electrons are highly anisotropic: across all energies, SEEs arrive first in the SOUTH (blue lines) and SUN (black lines) detectors before arrival in their NORTH (green) or ASUN (red) counterparts.
Compared to the observations along the Sun-Earth line by Wind and THEMIS-ARTEMIS, the dispersion in the normalized fluxes as a function of energy at STEREO-A is less pronounced, with the value of $\Delta t$ remaining near $\sim20$ min across all energy channels.

For all four spacecraft, we estimate the distance traveled by the SEEs after their initial detection traveling along the local magnetic field, before they returned to the spacecraft and were detected traveling anti-parallel to the local magnetic field.
This ``round-trip'' distance $\Delta x$ covered during the time $\Delta t$ is shown in Table \ref{tab:paths} for select electron energies observed by the spacecraft.
Using the thresholds described above, the range in values for this round-trip distance spans from 1.2 AU up to values of 2.8 AU, with an average value for P1, P2, Wind, and STA of 1.75 AU, 1.50 AU, 2.38 AU, and 1.48 AU, respectively.

\begin{figure*}[t!]
    \centering
    \includegraphics[width=0.95\textwidth]{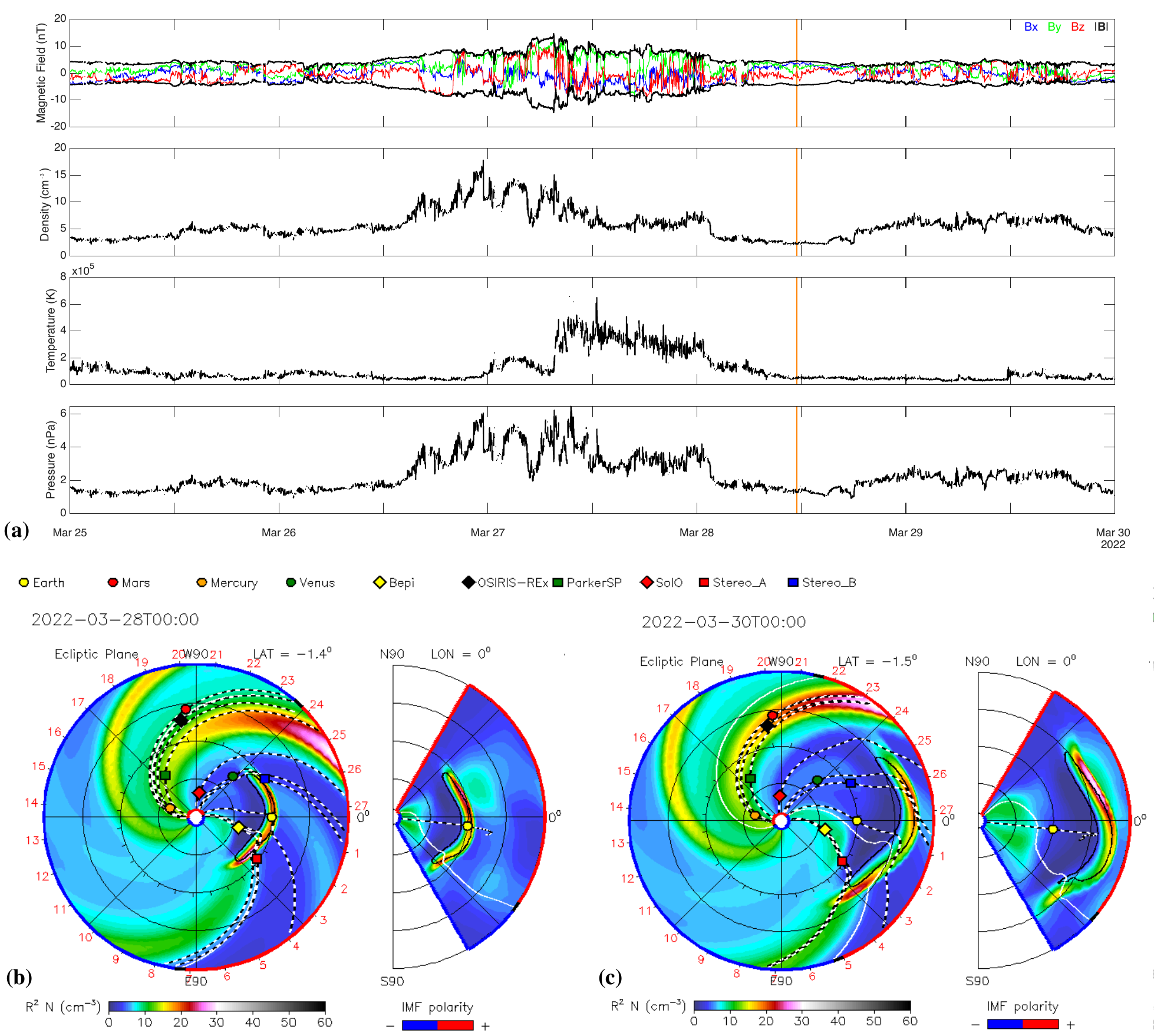}
    \caption{Solar wind properties from 25 March 2022 through 30 March 2022.
    (a) OMNI GSE magnetic field, proton density, temperature, and pressure.
    The orange vertical line denotes the time of the M4.0 flare responsible for the 28 March 2022 SEE event.
    (b--c) Snapshot of WSA-ENLIL+Cone model results for midnight UTC on (b) 28 March 2022 and (c) 30 March 2022.}
    \label{fig:omni}
\end{figure*}

If we assume that the bi-directional electron populations were injected at the same time and generated by the same mechanism, a round-trip value of $\Delta x < 2$ AU (for three out of the four spacecraft) suggests that the returning anti-field-aligned electrons could not have been generated near the Sun.
If instead these SEEs were reflected, then there must have been a magnetic boundary off of which these particles mirrored located a distance of $\frac{1}{2} \Delta x$ from the spacecraft, namely, at approximately 0.75 AU--1.2 AU.
To investigate this possibility, Figure \ref{fig:omni}a displays the solar wind magnetic field, proton density, temperature, and pressure from the OMNI dataset \citep{OMNI} over six days in March 2022.
The enhanced plasma density, temperature, and pressure after 12:00 UTC on 26 March visible in panel \ref{fig:omni}a illustrate that the ambient plasma environment was highly perturbed approximately 2 days prior to the onset of the SEE event.
The observations suggest that the solar wind at that time was influenced by a combination of a stream interaction region and a passing ICME, likely caused by the interaction between a high-speed stream and preceding transient ejecta.
This is also consistent with coronagraph observations from the \textit{SOlar and Heliospheric Observatory} (SOHO) spacecraft of a halo CME that erupted from the Sun near 06:00 UTC on 25 March 2022 \citep{SOHO}.
By 12:00 UTC on 28 March, two days after the ICME arrival, the solar wind properties had returned to their background, pre-perturbed state (after which the SEEs from the M4.0 flare were detected at Earth; see, e.g., Figure \ref{fig:fluxes2022}).
In addition to the OMNI data, Figures \ref{fig:omni}b and \ref{fig:omni}c display two snapshots in time from a run of the WSA-ENLIL+Cone model \citep{Odstrcil2003,Arge2004} available at the the NASA Goddard Space Flight Center Community Coordinated Modeling Center's Database Of Notifications, Knowledge, Information (CCMC DONKI).
Panel \ref{fig:omni}b illustrates that this model estimated an impact of an ICME near midnight on 28 March, approximately 36 h after the true arrival time observed in panel \ref{fig:omni}a.
This represents a significant deviation from the observations, but not unreasonable, since such behavior is consistent with the findings of \cite{Riley2018} in the spread of the modeled arrival times.
These authors found that for WSA-ENLIL+Cone runs hosted at CCMC for events between 2013 and 2018, the average accuracy in the modeled onset time was on the order of $\pm10$ h, but with errors often exceeding $\pm40$ h depending on the event.
The large difference for this specific ICME suggests that it was moving at a substantially faster speed than estimated by the model.

Despite this discrepancy, the model and observations show that an ICME was located \textit{somewhere} beyond 1 AU during the onset of this SEE event.
However, to refine this distance, we assume that the WSA-ENLIL+Cone model predicted the location of this ICME accurately but with a 36 h time-delay.
Figure \ref{fig:omni}c therefore shows the model output 36 h {after} the initial onset of the SEEs detected by Wind, THEMIS-ARTEMIS, and STEREO-A (i.e., at midnight on 30 March).
As visible in panel \ref{fig:omni}c, the modeled ICME shock front was located approximately 0.75 AU away from Earth along the Sun-Earth line.
This is consistent with the values of $\Delta x$ presented in Table \ref{tab:paths}, especially for the THEMIS-ARTEMIS spacecraft, which suggest the SEEs traveled distances of 0.75 AU $\le \Delta x \le$ 0.875 AU away from the spacecraft before being reflected and returning.
In addition, the model suggests that STEREO-A was located closer to the shock front compared to the Earth-based assets at this time, at a radial separation of only $\sim0.5$ AU.
This reduced distance is likewise consistent with the estimates of $\Delta x$ provided in Table \ref{tab:paths} for STEREO-A from the bi-directional SEE signatures, which range from 0.6 AU $\le \Delta x \le$ 0.9 AU.
The cause of this azimuthal asymmetry in distance is the enhanced plasma density associated with a corotating interaction region near 90$^\circ$ E longitude visible in the bottom half of Figure \ref{fig:omni}b, which would locally slow the radial expansion of the ICME compared to the relatively reduced densities near 0$^\circ$ longitude visible in the right half of Figure \ref{fig:omni}b.
We note that if THEMIS-ARTEMIS and Wind were on magnetic field lines connected to this feature, pileup associated with the more slowly expanding portion of the ICME could also be a candidate for the electrons' mirror points, which would suggest that the distance $\Delta x$ was not purely radial but may have had an azimuthal component as well.

\begin{table}[t!]
    \caption{
          Measurements of dispersive electrons from THEMIS-ARTEMIS P1 and P2, Wind, and STEREO-A for the 28 March 2022 event.
          Energies $E$ denote the center energy of select instrument channels, with the corresponding velocity $v$ of an electron at that energy.
          The time difference between detection of anti-Sunward and Sunward traveling electrons is given by $\Delta t$ (see text for an exact definition), with the total, round-trip distance traveled given by $\Delta x$ and the average for each spacecraft given by $\overline{\Delta x}$. 
          Differences in displayed values are due to rounding.}
    \centering
    \begin{tabular}{l|c|c|c|c|c}
        & $E$   &  $v$  & $\Delta t$ & $\Delta x$  & $\overline{\Delta x}$  \\ 
        & (keV)  &  (AU/min)   &   (min)         &     (AU)       &    (AU)  \\ \hline
\textbf{P1} & & & &                   & 1.75 \\ 
           & 204 & 0.084  & 15  & 1.3 & \\
           &  93 & 0.064  & 33  & 2.1 & \\
           &  52 & 0.050  & 35  & 1.7 & \\
           &  31 & 0.040  & 48  & 1.9 & \\ \hline
\textbf{P2} & & & &                   & 1.50 \\ 
           & 204 & 0.084  & 12  & 1.0 & \\
           &  93 & 0.064  & 27  & 1.7 & \\
           &  52 & 0.050  & 34  & 1.7 & \\
           &  31 & 0.040  & 39  & 1.6 & \\ \hline
\textbf{Wind}       & & & &           & 2.38 \\
           & 310 & 0.094  & 21  & 2.0   &\\ 
           & 182 & 0.081  & 33  & 2.7   &\\ 
           & 108 & 0.068  & 39  & 2.6   &\\ 
           &  66 & 0.056  & 50  & 2.8   &\\ 
           &  40 & 0.045  & 55  & 2.5   &\\ 
           &  27 & 0.038  & 45  & 1.7   &\\ \hline
\textbf{STA} & & & &                  & 1.48 \\ 
           & 314 & 0.094  & 17  & 1.6   &\\ 
           & 240 & 0.088  & 15  & 1.3   &\\ 
           & 179 & 0.081  & 22  & 1.8   &\\ 
           & 135 & 0.073  & 21  & 1.6   &\\ 
           &  94 & 0.064  & 21  & 1.4   &\\ 
           &  70 & 0.057  & 21  & 1.2  & 
    \end{tabular}
    \label{tab:paths}
\end{table}

Despite the similarity between the expected location of the ICME and the calculated distances traveled by the electrons, we caution that the exact value of $\Delta x$ is strongly dependent on our defined start and end times used to calculate $\Delta t$.
Here, we have required the normalized flux of the initial field-aligned population to exceed a value of 0.8 ($t_\mathrm{start}$), and the normalized flux of the anti-field-aligned population to exceed 0.4 ($t_\mathrm{end}$).
But, the choice of these values directly affects the calculated distance of these particles before they return to the spacecraft.
Since the initial field-aligned electron population is highly beamed with a normalized flux that rapidly exceeds a value of 0.8, the distances are less sensitive to changes in the chosen onset time $t_\mathrm{start}$; $t_\mathrm{end}$ is therefore most strongly affected.
To estimate the degree to which these chosen values affect the estimated distance, we can instead define $t_\mathrm{end}$ as the time when the normalized flux of the anti-field-aligned electrons exceeds a value of 0.2.
This decrease of our threshold by a factor of 2 would reduce the value of $\Delta t$ for Wind and P2 by an average of approximately 15 minutes across all energies.
As a result, this implies the reflecting boundary is located approximately $0.5$ AU closer to the spacecraft compared to using a threshold of 0.4 (cf.\ Table \ref{tab:paths}).
The resulting estimated distance of the ICME (located $\sim 0.7$ AU from Wind) is then nearly identical to the estimates from the WSA-ENLIL+Cone model results in Figure \ref{fig:omni}c.

In conclusion, the findings presented for this solar energetic electron event on 28 March 2022 include a strong anisotropy in the distribution of the SEEs, similarities between the spectra from the two counter-streaming electron beams, evidence for the presence of a reflecting boundary located beyond the spacecraft associated with an earlier ICME, and the time-delay in the arrival of the SEEs resulting in a distance traveled that is consistent with the expected location of the ICME at the onset of the SEE event.
This event was widespread throughout the inner heliosphere with similar signatures observed by Wind and THEMIS-ARTEMIS, located along the Sun-Earth line, and by STEREO-A, located approximately $30^\circ$ off of this line.
Taken in combination, this evidence strongly suggests that observed field-aligned solar energetic electrons were initially accelerated by a flare and streamed outward from the Sun past the spacecraft, reflected off of an ICME that passed nearly 2 days prior, and then returned, detected as a collimated beam of anti-field-aligned electrons and resulting in clear bi-directional distributions observed by the spacecraft.

\subsection{The 17 May 2012 Event}
\begin{figure*}[p]
    \centering
    \includegraphics[width=0.73\textwidth]{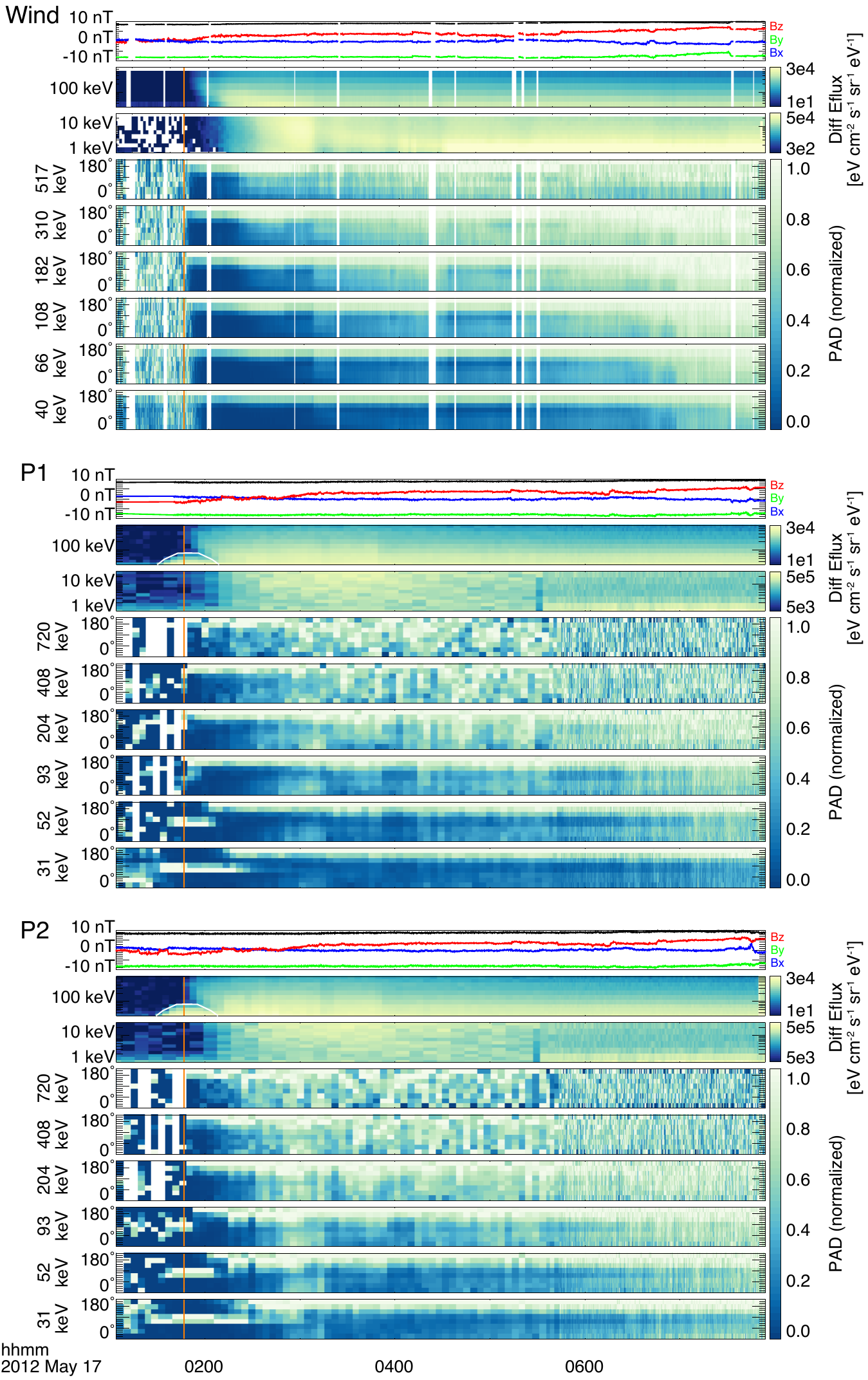}
    \caption{
        Observations of a solar energetic electron event on 17 May 2012 by Wind, THEMIS-ARTEMIS P1, and P2.
        In addition to the magnetic field and differential energy fluxes, the normalized electron pitch angle distributions from the SST instruments at select energies are shown.
        As in Figure \ref{fig:fluxes2022}, the vertical orange line denotes the time of the flare's peak X-ray flux, and the white semi-circles in the P1 and P2 SST channels denote contamination in the sensors from the flare X-rays.
        }
    \label{fig:2012}
\end{figure*}

Figure \ref{fig:2012} displays observations from Wind and the two THEMIS-ARTEMIS probes (P1 and P2, located in the solar wind at the time) during a SEE event that occurred on 17 May 2012.
All three spacecraft detect the first SEEs approximately 10--15 minutes after the M5.1 solar flare that peaked at 01:47 UTC.
The signatures were dispersive in time, with the highest energy electrons arriving to the spacecraft before electrons at lower energies.
Unlike the previous event, any IVD signature at higher energies in the electron differential flux is less convincing, with possibly only a hint in the Wind observations, likely caused by a different acceleration mechanism or connection to the flare region during this event.
Signatures of penetrating X-rays from the M-class flare are visible in the P1 and P2 omnidirectional fluxes (white semi-circle in Figure \ref{fig:2012}) which again do not represent physical electron fluxes.
This contamination is limited to electron energies below $E \lesssim 50$ keV.

\begin{sidewaysfigure*}[!p]
    \centering
    \includegraphics[width=\textwidth]{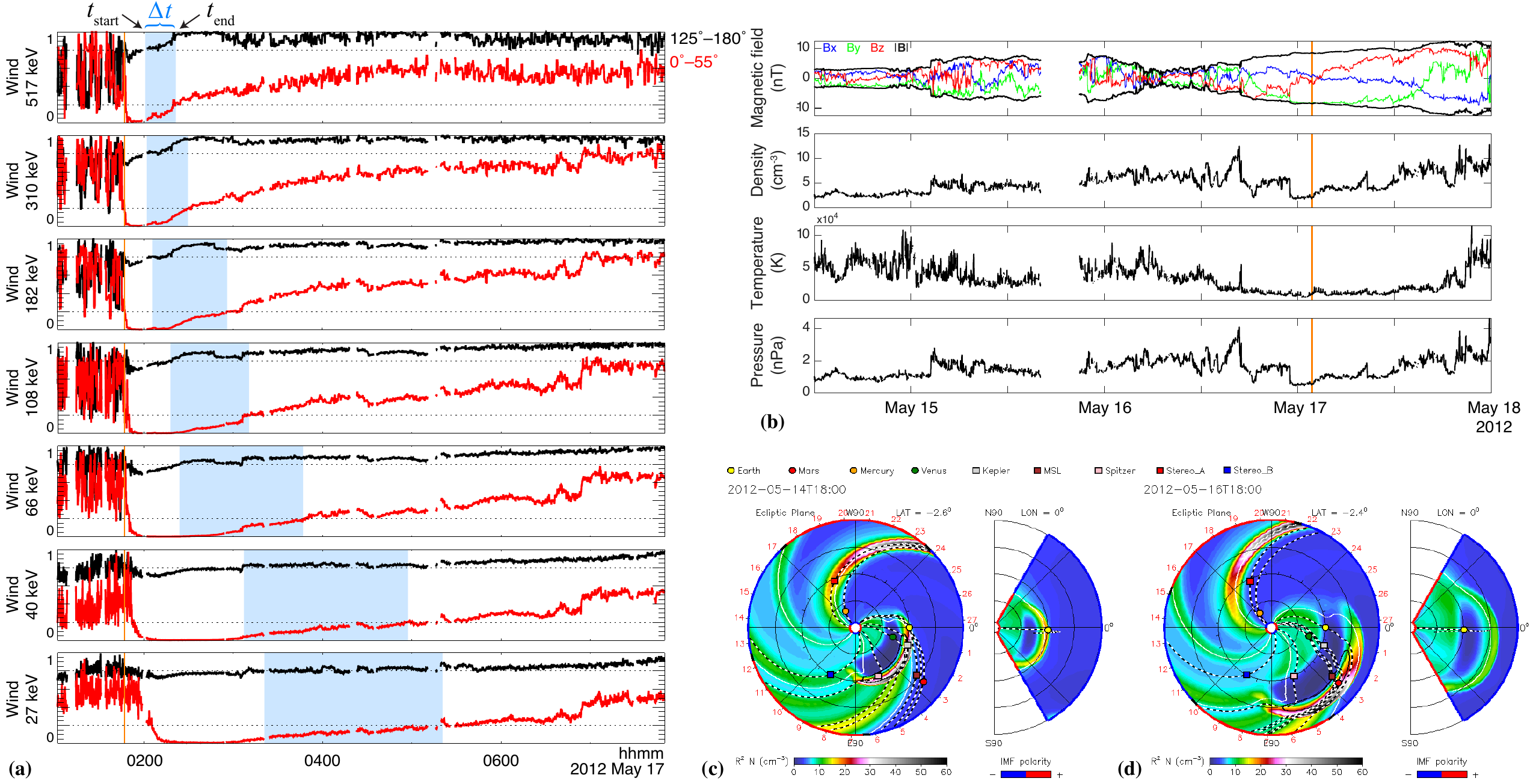}
    \caption{
            Properties of energetic electrons and solar wind conditions during the SEE event on 17 May 2012.
            (a) Timeseries of normalized electron fluxes at select energies observed by Wind, in the same style as Figure \ref{fig:windartlines2022}.
            Black or red lines correspond to electrons with anti-field-aligned or field-aligned pitch angles, respectively.
            (b) OMNI GSE magnetic field, proton density, temperature, and pressure in the same style as Figure \ref{fig:omni}.
            (c--d) Snapshots of WSA-ENLIL+Cone model results at 18:00 UTC on (c) 14 May 2012 and (c) 16 May 2012.
        }
    \label{fig:WindOmniWSA}
\end{sidewaysfigure*}

In addition to the magnetic field and omnidirectional differential energy fluxes, each grouping in Figure \ref{fig:2012} includes the energetic electron PADs from the SST instruments at select energies.
In the PAD for the lowest energy electrons, the penetrating X-rays are again visible, centered around the peak of the flare (vertical orange lines in the figure) and confined near pitch angles near $90^\circ$.
The ambient magnetic field during this event was predominately perpendicular to the Parker spiral direction with $B_z > 0$, with the spacecraft likely embedded within the remnants of an ICME (see Figure \ref{fig:WindOmniWSA} and discussion below).
As visible in Figure \ref{fig:2012}, the SEEs initially observed by Wind, P1, and P2 were strongly beamed, with nearly 100\% of the flux traveling anti-Sunward along the anti-field aligned direction.
Depending on electron energy, a second population of counter-streaming (i.e., field-aligned) electrons was detected anywhere from $\sim15$ min to 2 h later, similar to the 28 March 2022 event.
However in contrast to the other event, the counter-streaming electrons are more tenuous and do not exceed a normalized flux value above 0.4 until multiple hours after the first electrons were detected.

Figure \ref{fig:WindOmniWSA} shows properties of the solar energetic electrons as well as the ambient solar wind conditions during the SEE event in May 2012.
Panel \ref{fig:WindOmniWSA}a displays the timeseries of normalized electron fluxes observed by the Wind SST on 17 May 2012 during the event, in the same style as Figure \ref{fig:windartlines2022}.
Note that compared to the Figure \ref{fig:windartlines2022}, the coloring in panel \ref{fig:WindOmniWSA}a is reversed: although the population indicated by the black lines were still detected first by Wind, these electrons were anti-field-aligned during this event, while the red lines denote field-aligned electrons.
Since the normalized flux of the field-aligned electron population never exceeds a value of 0.4 until multiple hours after the initial detection of anti-field-aligned electrons, the blue shaded regions indicate where the ten-minute running average of the normalized flux of the anti-field-aligned population exceeds a value of 0.8 but the field-aligned flux is below 0.2.
As during the 28 March 2022 event, the dispersive nature of this 17 May 2012 event is visible by the value of $t_\mathrm{start}$, $t_\mathrm{end}$, and $\Delta t$ each increasing as a function of decreasing energy.

In the same style as Figure \ref{fig:omni}, Figures \ref{fig:WindOmniWSA}b--d display the (b) OMNI dataset during the 3 days in May 2012 surrounding the SEE event, as well as (c--d) two snapshots from the WSA-ENLIL+Cone modeling of a halo ICME that occurred on 11 May 2012, as also detected in SOHO coronagraph imaging.
Results from the WSA-ENLIL+Cone model, available at the CCMC DONKI website, suggest that the ICME would impact Earth near 18:00 UTC on 14 May 2012 (see Figure \ref{fig:WindOmniWSA}c).
However, Figure \ref{fig:WindOmniWSA}b shows an increase in the magnetic field magnitude, proton density, and pressure near 02:00 UTC on 15 May 2012, approximately 8 h later than the predicted ICME impact from the model.
Association of the passage of this ICME with these enhancements therefore suggests that the WSA-ENLIL+Cone estimated the arrival of the ICME approximately 8 h too early.
Hence, at the time of the SEE event onset, the shock front of this ICME would have been located at a distance of $\approx0.75$ AU beyond Earth, as illustrated in Figure \ref{fig:WindOmniWSA}d.

Notably, the interplanetary shock list from Richardson and Cane \citep{Richardson2010} does not include any shock impact until 16:00 UTC on 16 May 2012, visible in Figure \ref{fig:WindOmniWSA}b as a short-lived spike in proton density, pressure, and temperature, followed by a gradual increase of the IMF magnitude.
If instead these signatures denote the ICME's passage, then the estimated WSA-ENLIL+Cone impact time was nearly 46 h early.
This is still within the window found by \cite{Riley2018} but significantly larger than the average modeled error of $\pm10$ h, although this event was excluded from that analysis which only focused on ICMEs between 2013 and 2018.
It is also possible that since this specific simulation hosted at the CCMC used real-time observations of the initial properties of the CME, it did not accurately capture its evolution, resulting in this discrepancy.
Even so, a difference of 46 h suggests that the ICME would have been located \textit{at least} 1 AU beyond Earth at the beginning of the SEE event.
Unfortunately, the WSA-ENLIL+Cone simulation domain of the CCMC DONKI runs is limited to a size of 2 AU, and the modeled ICME exited this boundary around 12:00 on 18 May 2012
(i.e., only 34 h after SEEs were detected).
It is therefore not possible to further refine the location at the onset of the SEE event using this simulation.

Using the shading of $\Delta t$ in Figure \ref{fig:WindOmniWSA} and assuming the field-aligned population were reflected off of this passing ICME, the location where these electrons reflected ranges from 1.0 AU to 2.5 AU from the Wind spacecraft, with a mean distance across all energies of 1.9 AU.
The difference between the predicted location of the ICME from the model and observations is larger for this SEE event compared to the event in March 2022, highlighting the uncertainties associated with this ICME.

\section{Discussion and Conclusions}\label{sec:discuss}
We have presented two case studies of impulsive solar energetic electron events observed by multiple spacecraft throughout the solar system.
The first event we analyzed occurred on 28 March 2022, was associated with an M4.0 solar flare, and created a widespread SEE event that was detected by Wind and THEMIS-ARTEMIS located along the Sun-Earth line, as well as by STEREO-A which was $\sim30^\circ$ offset in azimuth.
The second SEE event on which we focused was also generated by an M5.1 flare and detected by Wind and THEMIS-ARTEMIS on 17 May 2012.
For each of these events, the associated energetic electron population was initially strongly collimated, traveling along the local magnetic field in a direction consistent with their origination from a flare located at the Sun.
These electrons were dispersive with energy during the May 2012 SEE event, but for the March 2022 event, displayed inverse velocity dispersion signature at higher energies.
Although IVD signatures have been detected for energetic ions inward of Earth's orbit, this represents the first such observation for energetic electrons at distances of 1 AU.

During both events, a second, counter-streaming population of electrons was later detected minutes to hours after the initial onset of energetic electrons, with arrival times that were also dispersive in time as a function of the electrons' energies.
At each energy, the counter-streaming population had a normalized flux that was $0.4$ to $0.8$ times the value of the initially detected electrons (for the May 2012 and March 2022 events, respectively).
The shape of the spectra for these two bi-directional electrons were nearly identical, indicating that these electrons were part of the same population.
Using the dispersion in the counter-streaming electrons' arrival times as a function of their energy, we have estimated that they traveled round-trip distances on the order of 1 AU between subsequent detections.
Finally, we have identified ICMEs that passed up to a few days prior to the SEE events' onsets that were plausibly located at the appropriate distances to form a boundary off of which the energetic electrons reflected in order to return to the spacecraft.

Similar bi-directional PADs for energetic electrons have been previously observed by Wind.
\cite{Wang2011} presented two SEE events with strong bi-directionality associated with outward-traveling electrons generated by a flare, and returning electrons that have reflected off of a boundary located beyond Earth.
However, the PADs were highly anisotropic only for energies below $\sim 40$ keV, and the normalized flux of the returning electrons remained below 0.1 times the flux of the outward-propagating electrons.
For higher energies up to 300 keV, \cite{Wang2011} found that the reflected electrons were less beamed, but with fluxes that reached nearly 0.8 times the outward-traveling electrons.
This is in contrast to the two events presented here, which were strongly beamed in both of the field-aligned and anti-field-aligned directions across all energies, with more comparable fluxes, suggesting that the propagation of these electrons was nearly scatter-free.

Bi-directional electron signatures are not observed during most solar energetic particle events.
To estimate their frequency, we have searched the fourteen years of THEMIS-ARTEMIS data for additional SEE events that display similar counter-streaming electron populations.
In order to avoid any contamination of the electron distributions by the terrestrial magnetosphere, we have restricted our search to SEE events that occurred when the probes were in the solar wind.
In addition, we also required that no solar energetic ions were present during the observations, since the electron sensors of the THEMIS-ARTEMIS SSTs can be easily contaminated by cross-talk from these particles.
After doing so, we have identified on the order of ten SEE events observed by THEMIS-ARTEMIS that display similarly clear, bi-directional signatures in the electron PADs.
With the restrictive conditions described above, this represents an underestimation for the occurrence rate for bi-directional energetic electron signatures, but still suggests their rarity compared to the hundreds of solar energetic particle events that have been detected by THEMIS-ARTEMIS.
We have not conducted a comprehensive search of the complete Wind or STEREO datasets to further refine this estimate.

Finally, bi-directional signatures in the distributions of solar energetic particles also occur within Earth's magnetotail near the Moon's orbit.
Previous studies have identified mechanisms by which energetic particles enter the tail far downstream of Earth, travel Earthward past the Moon, and then return after mirroring in the strong magnetic field closer to Earth \citep{Liuzzo2023,Liuzzo2024SEP}.
However, the occurrence of these counter-streaming particles \textit{within the solar wind} has implications for the amount of irradiation associated with energetic particle events, especially in the context of crewed exploration or long-term astronaut establishments located beyond low-Earth orbit.
Various studies have presented techniques to estimate the occurrence and/or severity of energetic particle events based on, e.g., the size and location of an active solar flare or the spectral properties of initial relativistic electrons detected in-situ (e.g., \citealp{Posner2007,Laurenza2009,Nunez2011,Torres2025}).
If successfully applied, these techniques are useful for providing advanced warning to astronauts or other assets on the lunar surface in the event of extreme solar particle irradiation.
However, for the two SEE events presented here, the initial onset of energetic particles was detected after a previous ICME passed the spacecraft, and for the 28 March 2022 event, the ambient solar wind conditions had even returned to their nominal, pre-ICME levels (see Figure \ref{fig:omni}).
Such a mechanism is not taken into account in current predictive capabilities, despite the clear influence on properties of the incident SEEs.
Our results therefore highlight the importance of characterizing the influence of \textit{preceding} space weather activity on properties of solar energetic particle events, which is therefore equally important for interpreting current conditions and predicting the impact of future radiation events.

\begin{acknowledgments}
L.L. and A.R.P. acknowledge support from NASA Lunar Data Analysis Program grant 80NSSC23K1338.
W.W. and C.O.L. acknowledge support from NASA grants 80NSSC21K1325, 80NSSC25K7690 and NSF grant AGS-2501437.
The authors acknowledge NASA contract NAS5-02099 for the use of THEMIS-ARTEMIS data, and specifically thank D. Larson and the late R.P. Lin for use of SST data, J.P. McFadden and the late C.W. Carlson for use of THEMIS-ARTEMIS ESA data, and K.-H. Glassmeier, U. Auster, and W. Baumjohann for the use of THEMIS-ARTEMIS FGM data provided under the lead of the Technical University of Braunschweig with financial support through the German Ministry for Economy and Technology and the German Center for Aviation and Space, Contract 50-OC-0302.
The authors also acknowledge the CCMC at Goddard Space Flight Center for the use of their DONKI tool, \href{https://kauai.ccmc.gsfc.nasa.gov/DONKI/}{kauai.ccmc.gsfc.nasa.gov/DONKI}.
L.L. is appreciative for discussions with LB Wilson III regarding the analysis of Wind data.

All data used in this study are openly available to the public: THEMIS-ARTEMIS data are located at \href{http://sprg.ssl.berkeley.edu/data/themis}{sprg.ssl.berkeley.edu/data/themis}, Wind data are located at \href{http://sprg.ssl.berkeley.edu/data/wind}{sprg.ssl.berkeley.edu/data/wind}, STEREO data are hosted at the STEREO Science Center (\href{https://stereo-ssc.nascom.nasa.gov/ins_data.shtml}{stereo-ssc.nascom.nasa.gov/ins\_data.shtml}), and OMNI data can be accessed through the Goddard Space Flight Center's Coordinated Data Analysis Web tool (\href{https://cdaweb.gsfc.nasa.gov}{cdaweb.gsfc.nasa.gov}).
WSA-ENLIL+Cone runs presented in this study are hosted by the CCMC, available at \href{https://kauai.ccmc.gsfc.nasa.gov/DONKI/view/WSA-ENLIL/19537/1}{kauai.ccmc.gsfc.nasa.gov/DONKI/view/WSA-ENLIL/19537/1} for the March 2022 event and \href{https://kauai.ccmc.gsfc.nasa.gov/DONKI/view/WSA-ENLIL/2381/1}{kauai.ccmc.gsfc.nasa.gov/DONKI/view/WSA-ENLIL/2381/1} for the May 2012 event.
\end{acknowledgments}

\begin{contribution}
L.L. was responsible for the formal analysis of Wind, THEMIS-ARTEMIS, and OMNI data, securing part of the funding supporting this study, writing the original version of the manuscript, as well as editing and submitting the manuscript.
W.W. provided STEREO data, and W.W. and C.O.L. contributed to its analysis.
L.L. and A.R.P. were responsible for the initial research concept.
All authors contributed equally in the discussions regarding the manuscript and in editing the manuscript.
\end{contribution}

\software{Space Physics Environment Data Analysis Software \citep{SPEDAS}, 
          Wind analysis software \citep{Wilson2021},
          SolarMACH \citep{SolarMACH}}


\begin{thebibliography}{}
\expandafter\ifx\csname natexlab\endcsname\relax\def\natexlab#1{#1}\fi
\providecommand{\url}[1]{\href{#1}{#1}}
\providecommand{\dodoi}[1]{doi:~\href{http://doi.org/#1}{\nolinkurl{#1}}}
\providecommand{\doeprint}[1]{\href{http://ascl.net/#1}{\nolinkurl{http://ascl.net/#1}}}
\providecommand{\doarXiv}[1]{\href{https://arxiv.org/abs/#1}{\nolinkurl{https://arxiv.org/abs/#1}}}

\bibitem[{M.~H. Acu{\~{n}}a {et~al.}(1995)Acu{\~{n}}a, Ogilvie, Baker, Curtis, Fairfield, \& Mish}]{Acuna1995}
Acu{\~{n}}a, M.~H., Ogilvie, K.~W., Baker, D.~N., {et~al.} 1995, \bibinfo{title}{{The Global Geospace Science Program and its investigations},} Space Science Reviews, 71, 5, \dodoi{10.1007/BF00751323}

\bibitem[{V. Angelopoulos(2008)Angelopoulos}]{Angelopoulos2008}
Angelopoulos, V. 2008, \bibinfo{title}{{The THEMIS Mission},} Space Science Reviews, 141, 5, \dodoi{10.1007/s11214-008-9336-1}

\bibitem[{V. Angelopoulos(2011)Angelopoulos}]{Angelopoulos2011}
Angelopoulos, V. 2011, \bibinfo{title}{{The ARTEMIS Mission},} Space Science Reviews, 165, 3, \dodoi{10.1007/s11214-010-9687-2}

\bibitem[{V. Angelopoulos {et~al.}(2019)Angelopoulos, Cruce, Drozdov, Grimes, Hatzigeorgiu, King, Larson, Lewis, McTiernan, Roberts, Russell, Hori, Kasahara, Kumamoto, Matsuoka, Miyashita, Miyoshi, Shinohara, Teramoto, Faden, Halford, McCarthy, Millan, Sample, Smith, Woodger, Masson, Narock, Asamura, Chang, Chiang, Kazama, Keika, Matsuda, Segawa, Seki, Shoji, Tam, Umemura, Wang, Wang, Redmon, Rodriguez, Singer, Vandegriff, Abe, Nose, Shinbori, Tanaka, UeNo, Andersson, Dunn, Fowler, Halekas, Hara, Harada, Lee, Lillis, Mitchell, Argall, Bromund, Burch, Cohen, Galloy, Giles, Jaynes, Le~Contel, Oka, Phan, Walsh, Westlake, Wilder, Bale, Livi, Pulupa, Whittlesey, DeWolfe, Harter, Lucas, Auster, Bonnell, Cully, Donovan, Ergun, Frey, Jackel, Keiling, Korth, McFadden, Nishimura, Plaschke, Robert, Turner, Weygand, Candey, Johnson, Kovalick, Liu, McGuire, Breneman, Kersten, \& Schroeder}]{SPEDAS}
Angelopoulos, V., Cruce, P., Drozdov, A., {et~al.} 2019, \bibinfo{title}{{The Space Physics Environment Data Analysis System (SPEDAS)},} Space Science Reviews, 215, 9, \dodoi{10.1007/s11214-018-0576-4}

\bibitem[{C. Arge {et~al.}(2004)Arge, Luhmann, Odstrcil, Schrijver, \& Li}]{Arge2004}
Arge, C., Luhmann, J., Odstrcil, D., Schrijver, C., \& Li, Y. 2004, \bibinfo{title}{{Stream structure and coronal sources of the solar wind during the May 12th, 1997 CME},} Journal of Atmospheric and Solar-Terrestrial Physics, 66, 1295, \dodoi{10.1016/j.jastp.2004.03.018}

\bibitem[{C.~M.~S. Cohen {et~al.}(2024)Cohen, Leske, Christian, Cummings, de~Nolfo, Desai, Giacalone, Hill, Labrador, McComas, McNutt, Mewaldt, Mitchell, Mitchell, Muro, Rankin, Schwadron, Sharma, Shen, Szalay, Wiedenbeck, Xu, Romeo, Vourlidas, Bale, Pulupa, Kasper, Larson, Livi, \& Whittlesey}]{Cohen2024}
Cohen, C. M.~S., Leske, R.~A., Christian, E.~R., {et~al.} 2024, \bibinfo{title}{{Observations of the 2022 September 5 Solar Energetic Particle Event at 15 Solar Radii},} The Astrophysical Journal, 966, 148, \dodoi{10.3847/1538-4357/ad37f8}

\bibitem[{Z. Ding {et~al.}(2025{\natexlab{a}})Ding, Wimmer-Schweingruber, Kollhoff, K{\"{u}}hl, Yang, Berger, Kouloumvakos, Wijsen, Guo, Pacheco, Li, Temmer, Rodriguez-Pacheco, Allen, Ho, Mason, Xu, \& Gunaseelan}]{Ding2025a}
Ding, Z., Wimmer-Schweingruber, R.~F., Kollhoff, A., {et~al.} 2025{\natexlab{a}}, \bibinfo{title}{{Investigation of the inverse velocity dispersion in a solar energetic particle event observed by Solar Orbiter},} Astronomy {\&} Astrophysics, 696, A199, \dodoi{10.1051/0004-6361/202553806}

\bibitem[{Z. Ding {et~al.}(2025{\natexlab{b}})Ding, Wimmer-Schweingruber, Chen, Zhao, Kollhoff, K{\"{u}}hl, Yang, Berger, Heidrich-Meisner, Rodriguez-Pacheco, Ho, Mason, Li, Form{\'{a}}nek, \& Owen}]{Ding2025}
Ding, Z., Wimmer-Schweingruber, R.~F., Chen, Y., {et~al.} 2025{\natexlab{b}}, \bibinfo{title}{{Bidirectional anisotropic solar energetic particle events observed by Solar Orbiter},} Astronomy {\&} Astrophysics, 1, \dodoi{10.1051/0004-6361/202556098}

\bibitem[{V. Domingo {et~al.}(1995)Domingo, Fleck, \& Poland}]{SOHO}
Domingo, V., Fleck, B., \& Poland, A.~I. 1995, \bibinfo{title}{{SOHO: The Solar and Heliospheric Observatory},} Space Science Reviews, 72, 81, \dodoi{10.1007/BF00768758}

\bibitem[{N. Dresing {et~al.}(2014)Dresing, G{\'{o}}mez-Herrero, Heber, Klassen, Malandraki, Dr{\"{o}}ge, \& Kartavykh}]{Dresing2014}
Dresing, N., G{\'{o}}mez-Herrero, R., Heber, B., {et~al.} 2014, \bibinfo{title}{{Statistical survey of widely spread out solar electron events observed with STEREO and ACE with special attention to anisotropies},} Astronomy and Astrophysics, 567, 1, \dodoi{10.1051/0004-6361/201423789}

\bibitem[{J. Gieseler {et~al.}(2023)Gieseler, Dresing, Palmroos, Freiherr~von Forstner, Price, Vainio, Kouloumvakos, Rodr{\'{i}}guez-Garc{\'{i}}a, Trotta, G{\'{e}}not, Masson, Roth, \& Veronig}]{SolarMACH}
Gieseler, J., Dresing, N., Palmroos, C., {et~al.} 2023, \bibinfo{title}{{Solar-MACH: An open-source tool to analyze solar magnetic connection configurations},} Frontiers in Astronomy and Space Sciences, 9, \dodoi{10.3389/fspas.2022.1058810}

\bibitem[{N. Gopalswamy {et~al.}(2013)Gopalswamy, Xie, Akiyama, Yashiro, Usoskin, \& Davila}]{Gopalswamy2013}
Gopalswamy, N., Xie, H., Akiyama, S., {et~al.} 2013, \bibinfo{title}{{THE FIRST GROUND LEVEL ENHANCEMENT EVENT OF SOLAR CYCLE 24: DIRECT OBSERVATION OF SHOCK FORMATION AND PARTICLE RELEASE HEIGHTS},} The Astrophysical Journal, 765, L30, \dodoi{10.1088/2041-8205/765/2/L30}

\bibitem[{R. Harten \& K. Clark(1995)Harten \& Clark}]{Harten1995}
Harten, R., \& Clark, K. 1995, \bibinfo{title}{{The design features of the GGS wind and polar spacecraft},} Space Science Reviews, 71, 23, \dodoi{10.1007/BF00751324}

\bibitem[{M.~L. Kaiser {et~al.}(2008)Kaiser, Kucera, Davila, St.~Cyr, Guhathakurta, \& Christian}]{Kaiser2008}
Kaiser, M.~L., Kucera, T.~A., Davila, J.~M., {et~al.} 2008, \bibinfo{title}{{The STEREO Mission: An Introduction},} Space Science Reviews, 136, 5, \dodoi{10.1007/s11214-007-9277-0}

\bibitem[{S. Krucker {et~al.}(2007)Krucker, Kontar, Christe, \& Lin}]{Krucker2007}
Krucker, S., Kontar, E.~P., Christe, S., \& Lin, R.~P. 2007, \bibinfo{title}{{Solar Flare Electron Spectra at the Sun and near the Earth},} The Astrophysical Journal, 663, L109, \dodoi{10.1086/519373}

\bibitem[{S. Krucker {et~al.}(1999)Krucker, Larson, Lin, \& Thompson}]{Krucker1999}
Krucker, S., Larson, D.~E., Lin, R.~P., \& Thompson, B.~J. 1999, \bibinfo{title}{{On the Origin of Impulsive Electron Events Observed at 1 AU},} The Astrophysical Journal, 519, 864, \dodoi{10.1086/307415}

\bibitem[{D. Lario {et~al.}(2009)Lario, Aran, \& Decker}]{Lario2009}
Lario, D., Aran, A., \& Decker, R.~B. 2009, \bibinfo{title}{{Major Solar Energetic Particle Events of Solar Cycles 22 and 23: Intensities Close to the Streaming Limit},} Solar Physics, 260, 407, \dodoi{10.1007/s11207-009-9463-1}

\bibitem[{D.~E. Larson {et~al.}(2015)Larson, Lillis, Lee, Dunn, Hatch, Robinson, Glaser, Chen, Curtis, Tiu, Lin, Luhmann, \& Jakosky}]{Larson2015}
Larson, D.~E., Lillis, R.~J., Lee, C.~O., {et~al.} 2015, \bibinfo{title}{{The MAVEN Solar Energetic Particle Investigation},} Space Science Reviews, 195, 153, \dodoi{10.1007/s11214-015-0218-z}

\bibitem[{M. Laurenza {et~al.}(2009)Laurenza, Cliver, Hewitt, Storini, Ling, Balch, \& Kaiser}]{Laurenza2009}
Laurenza, M., Cliver, E.~W., Hewitt, J., {et~al.} 2009, \bibinfo{title}{{A technique for short‐term warning of solar energetic particle events based on flare location, flare size, and evidence of particle escape},} Space Weather, 7, \dodoi{10.1029/2007SW000379}

\bibitem[{G. Li {et~al.}(2020)Li, Wu, Zhao, \& Yao}]{Li2020}
Li, G., Wu, X., Zhao, L., \& Yao, S. 2020, \bibinfo{title}{{Observations of Outward-propagating and Mirroring of the Same Energetic Electrons by Wind},} The Astrophysical Journal Letters, 905, L1, \dodoi{10.3847/2041-8213/abca87}

\bibitem[{Y. Li {et~al.}(2025)Li, Guo, Pacheco, Wang, Temmer, Ding, \& Wimmer-Schweingruber}]{Li2025}
Li, Y., Guo, J., Pacheco, D., {et~al.} 2025, \bibinfo{title}{{The delayed arrival of faster solar energetic particles as a probe into the shock acceleration process},} National Science Review, 12, \dodoi{10.1093/nsr/nwaf348}

\bibitem[{R. Lin(1974)Lin}]{Lin1974}
Lin, R. 1974, \bibinfo{title}{{Non-relativistic solar electrons},} Space Science Reviews, 16, 189, \dodoi{10.1007/BF00240886}

\bibitem[{R.~P. Lin(1985)Lin}]{Lin1985}
Lin, R.~P. 1985, \bibinfo{title}{{Energetic solar electrons in the interplanetary medium},} Solar Physics, 100, 537, \dodoi{10.1007/BF00158444}

\bibitem[{R.~P. Lin {et~al.}(1995)Lin, Anderson, Ashford, Carlson, Curtis, Ergun, Larson, McFadden, McCarthy, Parks, Reme, Bosqued, Coutelier, Cotin, D'Uston, Wenzel, Sanderson, Henrion, Ronnet, \& Paschmann}]{Lin1995}
Lin, R.~P., Anderson, K.~A., Ashford, S., {et~al.} 1995, \bibinfo{title}{{A three-dimensional plasma and energetic particle investigation for the Wind spacecraft},} Space Science Reviews, 71, 125, \dodoi{10.1007/BF00751328}

\bibitem[{L. Liuzzo {et~al.}(2024)Liuzzo, Poppe, Lee, \& Angelopoulos}]{Liuzzo2024SEP}
Liuzzo, L., Poppe, A.~R., Lee, C.~O., \& Angelopoulos, V. 2024, \bibinfo{title}{{Solar Energetic Electron Access to the Moon Within the Terrestrial Magnetotail and Shadowing by the Lunar Surface},} Geophysical Research Letters, 51, \dodoi{10.1029/2024GL110228}

\bibitem[{L. Liuzzo {et~al.}(2023)Liuzzo, Poppe, Lee, Xu, \& Angelopoulos}]{Liuzzo2023}
Liuzzo, L., Poppe, A.~R., Lee, C.~O., Xu, S., \& Angelopoulos, V. 2023, \bibinfo{title}{{Unrestricted Solar Energetic Particle Access to the Moon While Within the Terrestrial Magnetotail},} Geophysical Research Letters, 50, 1, \dodoi{10.1029/2023GL103990}

\bibitem[{D.~J. McComas {et~al.}(2016)McComas, Alexander, Angold, Bale, Beebe, Birdwell, Boyle, Burgum, Burnham, Christian, Cook, Cooper, Cummings, Davis, Desai, Dickinson, Dirks, Do, Fox, Giacalone, Gold, Gurnee, Hayes, Hill, Kasper, Kecman, Klemic, Krimigis, Labrador, Layman, Leske, Livi, Matthaeus, McNutt, Mewaldt, Mitchell, Nelson, Parker, Rankin, Roelof, Schwadron, Seifert, Shuman, Stokes, Stone, Vandegriff, Velli, von Rosenvinge, Weidner, Wiedenbeck, \& Wilson}]{McComas2016}
McComas, D.~J., Alexander, N., Angold, N., {et~al.} 2016, \bibinfo{title}{{Integrated Science Investigation of the Sun (ISIS): Design of the Energetic Particle Investigation},} Space Science Reviews, 204, 187, \dodoi{10.1007/s11214-014-0059-1}

\bibitem[{J.~P. McFadden {et~al.}(2008)McFadden, Carlson, Larson, Ludlam, Abiad, Elliott, Turin, Marckwordt, \& Angelopoulos}]{McFadden2008a}
McFadden, J.~P., Carlson, C.~W., Larson, D., {et~al.} 2008, \bibinfo{title}{{The THEMIS ESA plasma instrument and in-flight calibration},} Space Science Reviews, 141, 277, \dodoi{10.1007/s11214-008-9440-2}

\bibitem[{L.~I. Miroshnichenko {et~al.}(2005)Miroshnichenko, Klein, Trottet, Lantos, Vashenyuk, Balabin, \& Gvozdevsky}]{Miroshnichenko2005}
Miroshnichenko, L.~I., Klein, K., Trottet, G., {et~al.} 2005, \bibinfo{title}{{Relativistic nucleon and electron production in the 2003 October 28 solar event},} Journal of Geophysical Research: Space Physics, 110, \dodoi{10.1029/2004JA010936}

\bibitem[{R. M{\"{u}}ller-Mellin {et~al.}(2008)M{\"{u}}ller-Mellin, B{\"{o}}ttcher, Falenski, Rode, Duvet, Sanderson, Butler, Johlander, \& Smit}]{STEREOSEPT}
M{\"{u}}ller-Mellin, R., B{\"{o}}ttcher, S., Falenski, J., {et~al.} 2008, \bibinfo{title}{{The Solar Electron and Proton Telescope for the STEREO Mission},} Space Science Reviews, 136, 363, \dodoi{10.1007/s11214-007-9204-4}

\bibitem[{M. N{\'{u}}{\~{n}}ez(2011)N{\'{u}}{\~{n}}ez}]{Nunez2011}
N{\'{u}}{\~{n}}ez, M. 2011, \bibinfo{title}{{Predicting solar energetic proton events (E {\&}gt; 10 MeV)},} Space Weather, 9, \dodoi{10.1029/2010SW000640}

\bibitem[{D. Odstrcil(2003)Odstrcil}]{Odstrcil2003}
Odstrcil, D. 2003, \bibinfo{title}{{Modeling 3-D solar wind structure},} Advances in Space Research, 32, 497, \dodoi{10.1016/S0273-1177(03)00332-6}

\bibitem[{N.~E. Papitashvili \& J.~H. King(2020)Papitashvili \& King}]{OMNI}
Papitashvili, N.~E., \& King, J.~H. 2020, \bibinfo{title}{{OMNI 1-min Data Set [Dataset]},} Space Physics Data Facility.
\newblock \url{https://spase-metadata.org/NASA/NumericalData/OMNI/HighResolutionObservations/Version1/PT1M}

\bibitem[{J. P{\'{e}}rez‐Peraza {et~al.}(2018)P{\'{e}}rez‐Peraza, M{\'{a}}rquez‐Adame, Miroshnichenko, \& Velasco‐Herrera}]{PerezPeraza2018}
P{\'{e}}rez‐Peraza, J., M{\'{a}}rquez‐Adame, J.~C., Miroshnichenko, L., \& Velasco‐Herrera, V. 2018, \bibinfo{title}{{Source Energy Spectrum of the 17 May 2012 GLE},} Journal of Geophysical Research: Space Physics, 123, 3262, \dodoi{10.1002/2017JA025030}

\bibitem[{A. Posner(2007)Posner}]{Posner2007}
Posner, A. 2007, \bibinfo{title}{{Up to 1‐hour forecasting of radiation hazards from solar energetic ion events with relativistic electrons},} Space Weather, 5, \dodoi{10.1029/2006SW000268}

\bibitem[{S. Purkhart {et~al.}(2024)Purkhart, Veronig, Kliem, Jarolim, Dissauer, Dickson, Podladchikova, \& Krucker}]{Purkhart2024}
Purkhart, S., Veronig, A.~M., Kliem, B., {et~al.} 2024, \bibinfo{title}{{Multipoint study of the rapid filament evolution during a confined C2 flare on 28 March 2022, leading to eruption},} Astronomy {\&} Astrophysics, 689, A259, \dodoi{10.1051/0004-6361/202450092}

\bibitem[{D. Reames(1999)Reames}]{Reames1999}
Reames, D. 1999, \bibinfo{title}{{Particle acceleration at the Sun and in the heliosphere},} Space Sci. Rev., 90, \dodoi{10.1023/A:1005105831781}

\bibitem[{D.~V. Reames(2021)Reames}]{Reames2021}
Reames, D.~V. 2021, Lecture Notes in Physics, Vol. 978, {Solar Energetic Particles} (Cham: Springer International Publishing), \dodoi{10.1007/978-3-030-66402-2}

\bibitem[{I.~G. Richardson \& H.~V. Cane(2010)Richardson \& Cane}]{Richardson2010}
Richardson, I.~G., \& Cane, H.~V. 2010, \bibinfo{title}{{Near-Earth Interplanetary Coronal Mass Ejections During Solar Cycle 23 (1996 - 2009): Catalog and Summary of Properties},} Solar Physics, 264, 189, \dodoi{10.1007/s11207-010-9568-6}

\bibitem[{P. Riley {et~al.}(2018)Riley, Mays, Andries, Amerstorfer, Biesecker, Delouille, Dumbovi{\'{c}}, Feng, Henley, Linker, M{\"{o}}stl, Nu{\~{n}}ez, Pizzo, Temmer, Tobiska, Verbeke, West, \& Zhao}]{Riley2018}
Riley, P., Mays, M.~L., Andries, J., {et~al.} 2018, \bibinfo{title}{{Forecasting the Arrival Time of Coronal Mass Ejections: Analysis of the CCMC CME Scoreboard},} Space Weather, 16, 1245, \dodoi{10.1029/2018SW001962}

\bibitem[{D. Ruffolo {et~al.}(2006)Ruffolo, Tooprakai, Rujiwarodom, Khumlumlert, Wechakama, Bieber, Evenson, \& Pyle}]{Ruffolo2006}
Ruffolo, D., Tooprakai, P., Rujiwarodom, M., {et~al.} 2006, \bibinfo{title}{{Relativistic Solar Protons on 1989 October 22: Injection and Transport along Both Legs of a Closed Interplanetary Magnetic Loop},} The Astrophysical Journal, 639, 1186, \dodoi{10.1086/499419}

\bibitem[{A. Saiz {et~al.}(2008)Saiz, Ruffolo, Bieber, Evenson, \& Pyle}]{Saiz2008}
Saiz, A., Ruffolo, D., Bieber, J.~W., Evenson, P., \& Pyle, R. 2008, \bibinfo{title}{{Anisotropy Signatures of Solar Energetic Particle Transport in a Closed Interplanetary Magnetic Loop},} The Astrophysical Journal, 672, 650, \dodoi{10.1086/523663}

\bibitem[{J.-A. Sauvaud {et~al.}(2008)Sauvaud, Larson, Aoustin, Curtis, M{\'{e}}dale, Fedorov, Rouzaud, Luhmann, Moreau, Schr{\"{o}}der, Louarn, Dandouras, \& Penou}]{STEREOSWEA}
Sauvaud, J.-A., Larson, D., Aoustin, C., {et~al.} 2008, \bibinfo{title}{{The IMPACT Solar Wind Electron Analyzer (SWEA)},} Space Science Reviews, 136, 227, \dodoi{10.1007/s11214-007-9174-6}

\bibitem[{L.~C. Tan {et~al.}(2007)Tan, Reames, \& Ng}]{Tan2007}
Tan, L.~C., Reames, D.~V., \& Ng, C.~K. 2007, \bibinfo{title}{{Bulk Flow Velocity and First‐Order Anisotropy of Solar Energetic Particles Observed on the Wind Spacecraft: Overview of Three “Gradual” Particle Events},} The Astrophysical Journal, 661, 1297, \dodoi{10.1086/516626}

\bibitem[{L.~C. Tan {et~al.}(2009)Tan, Reames, Ng, Saloniemi, \& Wang}]{Tan2009}
Tan, L.~C., Reames, D.~V., Ng, C.~K., Saloniemi, O., \& Wang, L. 2009, \bibinfo{title}{{OBSERVATIONAL EVIDENCE ON THE PRESENCE OF AN OUTER REFLECTING BOUNDARY IN SOLAR ENERGETIC PARTICLE EVENTS},} The Astrophysical Journal, 701, 1753, \dodoi{10.1088/0004-637X/701/2/1753}

\bibitem[{J. Torres {et~al.}(2025)Torres, Chan, Zhao, \& Zhang}]{Torres2025}
Torres, J., Chan, P.~K., Zhao, L., \& Zhang, M. 2025, \bibinfo{title}{{A Machine Learning Approach to Predicting SEP Proton Intensity and Events Using Time Series of Relativistic Electron Measurements},} Space Weather, 23, \dodoi{10.1029/2024SW003921}

\bibitem[{L. Wang {et~al.}(2011)Wang, Lin, \& Krucker}]{Wang2011}
Wang, L., Lin, R.~P., \& Krucker, S. 2011, \bibinfo{title}{{Pitch-angle distributions and temporal variations of 0.3-300 keV solar impulsive electron events},} Astrophysical Journal, 727, \dodoi{10.1088/0004-637X/727/2/121}

\bibitem[{W. Wei {et~al.}(2024)Wei, Lee, Dresing, Khoo, Jian, Luhmann, Cohen, Fraschetti, Zhuang, Huang, Owen, Nicolaou, Rodr{\'{i}}guez-Garc{\'{i}}a, Palmerio, Jebaraj, Lynch, \& Carcaboso}]{Wei2024}
Wei, W., Lee, C.~O., Dresing, N., {et~al.} 2024, \bibinfo{title}{{Very Large and Long-lasting Anisotropies Caused by Sunward Streaming Energetic Ions: Solar Orbiter and STEREO A Observations},} The Astrophysical Journal Letters, 973, L52, \dodoi{10.3847/2041-8213/ad78df}

\bibitem[{L.~B. Wilson(2021)Wilson}]{Wilson2021}
Wilson, L.~B. 2021, {lynnbwilsoniii/wind{\_}3dp{\_}pros: Space Plasma Missions IDL Software Library}, [Software]. Zenodo, \dodoi{10.5281/zenodo.4451330}

\end{thebibliography}

\bibliographystyle{aasjournalv7}



\end{document}